\documentclass[onecolumn,11pt]{article}

\usepackage{graphicx}    % needed for including graphics e.g. EPS, PS
\usepackage{color, soul}
\usepackage{amsmath, gensymb}
\usepackage{amssymb}
\usepackage{amsthm}
\usepackage{lineno}
\usepackage{graphicx}
\usepackage{epstopdf}
\usepackage{fixltx2e}
\usepackage{caption} % to add captions to figures\tables
\usepackage{float}  % to place figure at point
\usepackage{subcaption}
\usepackage{pdfpages}
\usepackage{import}
\usepackage{booktabs} % merge columns
\usepackage{wasysym} % to add weird symbols. remove when not needed.
\usepackage{txfonts}   % for wierd font in formulas
\usepackage{hyperref} % made to set a table of contents for a pdf
\usepackage{hypcap}
\usepackage{breakcites} % allow line breaks in citations
\usepackage{ulem} % for strikeout text
\usepackage{bm} % bold math symbols
\def\timur[#1]{\textcolor{blue}{#1}}
\def\igor[#1]{\textcolor{magenta}{#1}}

\begin{document}
%**************** TITLE ********************************************
\title{Embedded Fracture Model for Coupled Flow and Geomechanics}

\author{I. Shovkun and T. Garipov and H. A. Tchelepi}
\date{}
\maketitle

\begin{center}
Energy Resources Engineering, Stanford University, USA \end{center}

\textbf{Abstract:}
Fluid  injection and  production cause  changes in  reservoir pressure,  which
result  in deformations  in the  subsurface. This  phenomenon is  particularly
important in reservoirs with abundant fractures and faults because the induced
slip  and opening  of the  fractures may  significantly alter  their hydraulic
properties. Modeling  strongly coupled poro-mechanical processes  in naturally
fractured reservoirs  is a  challenging problem.
The  Discrete Fracture  Model  (DFM)  is a  state-of-art  method for  modeling
coupled  flow and  mechanics  in fractured  reservoirs.  This method  requires
constructing  computational grids  that comform  to fractures,  which is  very
challenging in complex 3D settings. The  objective of this study is to develop
a numerical  method that  does not  require gridding  near fractures  and can
efficiently model hydromechanical interactions in fractured reservoirs.
We utilize formulations  based on the Strong Discontinuity  Approach (SDA) for
mechanics  and Embedded  Discrete Fracture  Model  (EDFM) for  flow. We  first
present  a mathematical  formulation and  emphasize the  kinematic aspects  of
fracture slip and opening. We then introduce a series of mechanical tests that
investigate the spatial convergence of the model and compare its accuracy with
the Discrete  Fracture Model  (DFM). We finally  consider a  synthetic coupled
case of a reservoir with several  fractures and compare the performance of the
SDA and DFM methods. Our  results indicate super-linear spatial convergence of
the proposed SDA algorithm. Numerical simulations confirm the applicability of
the  proposed   method  to  modeling   the  coupling  effects   in  subsurface
applications.


%===================================================================
\section{Introduction}

% High-level introduction
%
Field operations in petroleum, geothermal, and waste disposal applications
frequently involve injecting and withdrawing fluids from the subsurface.
Changes in reservoir pressure in naturally fractured reservoirs can lead to
reactivation of natural fractures and faults
\cite{grasso1992mechanics,zoback2002production,rutledge2004faulting,zoback2012earthquake,hwang2015stress}.
Fault reactivation often  leads to shear of the  wellbore casing or damage
of the reservoir  integrity \cite{elf1992monitoring,wiprut2000fault}. In tight
rocks,  fracture  reactivation may  cause  significant  changes in  reservoir
permeability \cite{min2004stress}.

Modeling  fracture  reactivation  is  possible with  a  variety  of  numerical
methods. It has been shown that  the Finite Element Method (FEM) with discrete
fracture representation (DFM) is an accurate and efficient method for modeling
discontinuities                 in                porous                 media
\cite{cappa2011modeling,Fu2016,SALIMZADEH2018212}.   This   method,   however,
requires  the  usage  of   sophisticated  gridding  techniques  for  preparing
high-quality computational grids.
The Extended/Generalized  Finite Element (xFEM) and  Phase-Field methods allow
to  handle arbitrary  fracture geometries  and, therefore,  suit for  modeling
fracture  propagation   \cite{duarte2000generalized,heister2015primal}.  These
methods introduce additional global variables to improve the representation of
the  deformation field  \cite{haddad2015integration,shovkun2018geomechanical}.
This leads  to a significant  computational overhead compared  to conventional
FE-based methods.
Several  approaches  to  modeling  fractures  have  been  developed  based  on
non-local  flomulations  of  continuum mechanics.  Peridynamics  and  Discrete
Element methods  are among  the most capable  techniques for  modeling complex
geometrical                       fracture                      configurations
\cite{zhao2009numerical,ouchi2015fully,sun2016assessment}.  Similarly to  xFEM
and  Phase-Field,   however,  these   non-local  methods  have   very  high
computational cost.
The Boundary Element and Displacement Discontinuity methods are very efficient
techniques;  they are  widely applied  to mechanical  problems with  fractures
\cite{olson2008multi,sesetty2012simulation}.  These  methods  allow  to  model
reservoirs  with  high numbers  of  fractures;  however, they  cannot  capture
heterogeneous and  anisotropic rock properties  \cite{mcclure2012modeling}. It
has been  shown that  the Boundary  Element methods can  be combined  with the
Finite   Element   methods   to  solve   coupled   hydro-mechanical   problems
\cite{Norbeck2016embedded}.

% SDA
%
Another  type of  Finite  Element mechanical  models is  based  on the  Strong
Discontinuity Approach (SDA) \cite{oliver1996modelling}.  This method uses the
``embedded''  fracture representation  and can  be applied  to model  fracture
activation   and  propagation   \cite{oliver1999strong}.  The   method  shares
similarities with xFEM  but captures the deformation  of discontinuity locally
\cite{linder2007finite}. The nonlinear return  mapping algorithm allows SDA to
reduce the  number of degrees  of freedom to that  of a linear  elastic system
without fractures \cite{regueiro2001plane}.  This localization allows seamless
incorporation of  the method into  the existing Finite Element  software and
renders the  method very attractive  from the numerical  simulation standpoint
\cite{mosler2005novel}.

There  are several  numerical  techniques to  simulate  fluid flow  in
fractured   formations.   Dual   permeability   and   dual   porosity   models
\cite{warren1963behavior},  the  Discrete   Fracture  Network  approach  (DFN)
\cite{cacas1990modeling}, Discrete Fracture  Models \cite{KarimiFard2004}, and
the           Multiple          Interacting           Continua          (MINC)
\cite{narasimhan1988minc,pruess1992brief}  methods  are  widely  used  in  the
industry.
Several approaches based on local permeability modification have been proposed
to  model  flow  in  fracture  media  in  an  upscaled  dual-porosity  fashion
\cite{gong2008upscaling,li2015effective}. The application of the Finite Volume
method with the embedded fracture  representation (EDFM) has been investigated
for the past decade \cite{li2008efficient,chai2018uncertainty}.

% SDA coupled
%
While substantial  research exists on using  SDA in solid mechanics,  very few
studies  investigate  the  application   of  SDA  to  coupled  flow-mechanical
problems.  The  approach of  enriching  a  global  variable to  approximate  a
discontinuity used in  SDA was also applied to model  discontinuities in fluid
pressure  \cite{alfaiate2010use}.  The  authors,  however,  did  not  consider
hydraulically active  fractures with  flow within them.  A cobbination  of SDA
enrichments  for both  displacement and  flow discontinuities  can be  used to
model the  coupled behavior of fractured  media \cite{callari2002finite}. This
approach and the  aforementioned SDA model for  flow \cite{alfaiate2010use} do
not allow to model highly conductive fractures.

A  recent  advancement combined  a  mechanical  model with  embedded  fracture
representation  and a  conventional EDFM  for flow  \cite{deb2017modeling}. In
their work,  Deb and Jenny utilized  the Extended Finite Volume  method with a
stabilization term for  fracture activation. Their model,  however, is limited
to a 2D setting  and slipping fractures, and it treats the  fracture jump as an
independent variable. An approach that consists of a Finite Element mechanical
model  with  SDA and  an  embedded  fracture model  for  flow  would
allow to model reservoirs with conductive fractures efficiently.

% This paper
%
In this  manuscript, we  present a  new hybrid  numerical method  for modeling
coupled fluid flow and geomechanics  with embedded fracture representation. We
combine an  FEM-based mechanical SDA model  and an FVM-based flow  EDFM model,
which, to the best of our knowledge,  has not been done before. The mechanical
formulation    utilizes     SDA    with    the     return-mapping    algorithm
\cite{borja2013plasticity}.  This  paper  consists  of  comparison  cases
between   the  proposed   EDFM  model   and   the  DFM   model  described   in
\cite{garipov2016discrete}. We  first investigate  the spatial  convergence of
the  mechanical SDA  model  in plane  strain scenarios with single fractures.
Second, we investigate  the spatial convergence of the EDFM
flow  model on  a  single-phase  stationary problem.  We  finally compare  the
results produced by the EDFM and DFM  models in a coupled 3D case with several
fractures.

\section{Theory}
In this  section, we introduce  a system  of equations that  describes coupled
fluid flow and rock deformations in  a saturated porous medium with fractures.
We  employ linear  poroelasticity equations  and Darcy's  law to  describe the
single-phase  flow   and  mechanics   within  a  compressible   porous  medium
\cite{Coussy2004}.
We then outline the governing equations for  single-phase   flow  within
fractures and corresponding fractures deformations.

\subsection{Porous media}
\subsubsection{Fluid flow}
The mass conservation of a single-phase fluid in a porous medium follows
the continuity equation
\begin{equation} \label{eq:continuity}
  \frac{ \partial \left( \rho_f \phi \right) }{\partial t} =
  \nabla \cdot \left( \rho_f \bm{v} \right) + Q_{FM},
\end{equation}
where $\rho_f$ is the fluid density,
$\phi$ is the rock porosity,
$\bm{v}$ is the fluid superficial velocity,
$Q_{FM}$ is the mass flux from the fracture into the matrix,
and
$t$ is the time.
Assuming that in an isothermal system a sligtly compressible fluid
\begin{equation} \label{eq:fl-comp}
  \frac{d \rho_f}{d p_M} = c_f \rho_f,
\end{equation}
complies with the Darcy's law
\begin{equation} \label{eq:darcy}
  \bm{v} = - \frac{\bm{k}_M}{\mu_f} \left(\nabla p_M - \rho_f \bm{g} \right)
\end{equation}
and that the poroelastic solid is linear and isotropic
\begin{equation} \label{eq:matrix-comp}
  d\phi = b\ d\left(\nabla \cdot \bm{u}\right) + \frac{1}{K} \left( 1-b \right) \left( b - \phi \right) dp_M,
\end{equation}
%
% we modify Eq. \ref{eq:continuity} to arrive at \cite{Coussy2010}:
Eq. \eqref{eq:continuity} becomes \cite{Coussy2010}:
\begin{equation} \label{eq:matrix-flow}
  \frac{1}{M} \frac{\partial p_M}{\partial t} +
  b \frac{\partial \left( \nabla \cdot \bm{u} \right)}{\partial t} -
  \nabla \cdot \frac{\bm{k}_M}{\mu_f} \left( \nabla p_M - \rho_f \bm{g} \right)
  = q_{MF},
\end{equation}
where
% where $p_M$ is the fluid pressure in the rock matrix,
% $t$ is the time variable,
$\bm{u}$ is the solid matrix displacement,
$b$ is the rock Biot coefficient,
$\bm{k}_M$ is the matrix permeability,
$\mu_f$ is the fluid viscosity,
$\rho_f$ is the fluid mass density,
$M = \phi c_f + 1 / K \left( 1-b \right) \left( b - \phi \right)$ is the Biot modulus,
$K$ is the drained bulk modulus of the rock,
$\bm{g}$ is the gravity acceleration constant,
and $q_{MF} = Q_{MF}/\rho_F$ is the volumetric matrix-fracture flux.

\subsubsection{Mechanical equilibrium}
The  mechanical equilibrium  of the  system is  described by  the quasi-static
Cauchy equation:
\begin{equation} \label{eq:cauchy}
\nabla \cdot \textbf{S} = -\rho_b \bm g
\end{equation}
where $\textbf{S}$ is the total stress tensor and
$\rho_b$ is the bulk mass density of the saturated rock.
We use the following isotropic relation between total and effective stresses
\cite{Zoback2013}: % p. 68: sigma = S - alpha p -- reservoir geomechanics
\begin{equation} \label{eq:effective}
\textbf{S} = \bm \sigma + b p_M\textbf{I}
\end{equation}
where $\bm\sigma$ is the effective stress tensor, $p$ is pore pressure,
$b$ is the scalar Biot coefficient,
and $\textbf I$ is the second-order unit tensor. % just to keep things simple - no tensor
We further assume a linear relationship between the stress and strain tensors:
\begin{equation} \label{eq:stress-strain}
  \bm \sigma = \varmathbb{C} : \bm \varepsilon,
\end{equation}
where $\varmathbb{C}$ is the fourth-order linear elastic stiffness tensor.

\subsection{Fractures}
\subsubsection{Fluid flow}
The governing equation of the fluid flow  in fractures is very similar to that
in  the matrix.  We neglect,  however, the  coupling terms  that describe  the
changes in matrix aperture permeability due to pressure perturbations:
%

% An equation that describes single-phase fluid flow in a fracture is similar to
% Eq.  \ref{eq:matrix-flow}.  We  neglect,  however,  the  coupling  terms  that
% describe  the  changes  in  matrix   aperture  permeability  due  to  pressure
% perturbations. This  substantial assumption  will be  addressed in  our future
% work.
%
\begin{equation} \label{eq:fracture-flow}
  c_f \frac{\partial p_F}{\partial t} -
  \nabla \cdot \frac{\bm{ k }_F}{\mu_f} \left( \nabla p_F - \rho_f \bm{g} \right)
  = q_{FF} - q_{MF}
\end{equation}
where $c_f$ is the fluid compressibility,
$p_F$ is the fluid pressure in the fracture,
$_F$ is the fracture permeability,
$q_{FF}$ is the flux between intersecting fractures.
% and $q_{WF}$ is the well-fracture source term. \timur[let's delete this term],
The computation of the terms $q_{FF}$ and $q_{MF}$ is discussed in Section
\ref{sec:frac-flow}.

\subsubsection{Contact mechanics} \label{sec:mechanics}
\begin{figure}[H]
  \centering
  \includegraphics[width=0.7\textwidth]{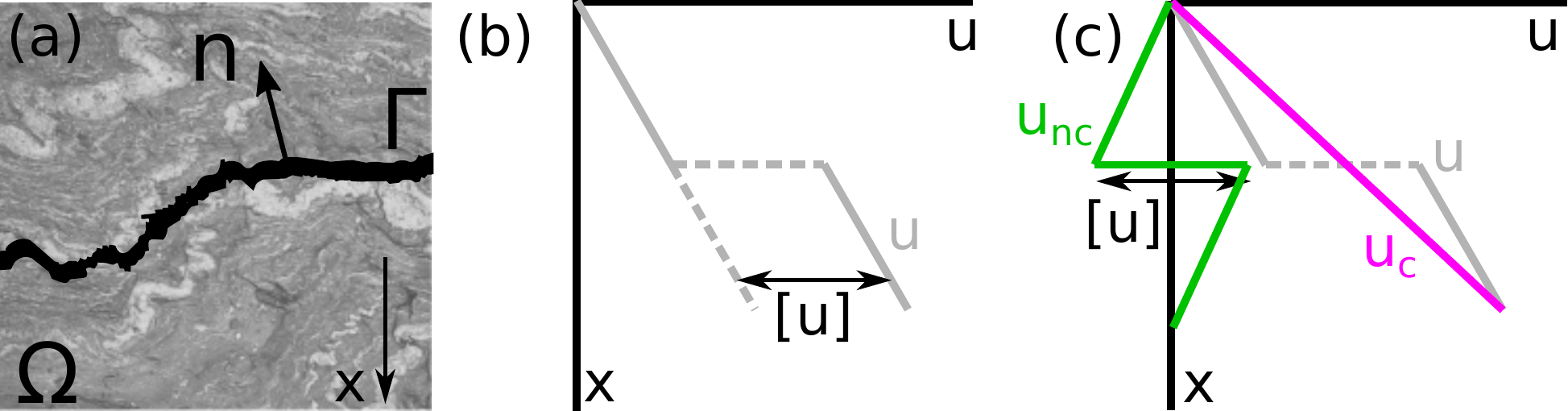}
  \caption{
    (a) A body $\Omega$ that contains a discontinuity $\Gamma$ with a normal $\bm{n}$.
    (b) Displacement field exhibits a jump across the fracture.
    (c) Decomposition of the displacement field into a conforming (regular) and
    non-conforming (singular) parts.
  }\label{fig:enrichment}
\end{figure}

\paragraph{Kinematics:}
We further consider  a domain $\Omega$ separated by  a two-dimensional surface
$\Gamma$  that  represents  a  fracture  (Fig.  \ref{fig:enrichment}a).  Since
fractures are discontinuities, the displacement  field $\bm{u}$ % in this domain
is  discontinuous  (Fig.~\ref{fig:enrichment}b)  and  can  be  decomposed  as
\cite{simo1994new}
\begin{equation} \label{eq:discontinuity}
  \bm{u} =  \bm{\overline{u}} + \bm{\left[ u \right]} H_\Gamma,
\end{equation}
where $\bm{\left[ u \right]}$ is the jump vector and
$H_\Gamma$ is the Heaviside step function that is equal to zero on one side of $\Gamma$ and
equal to one on the other side.
SDA modifies Eq.~\ref{eq:discontinuity} by adding and subtracting a continuous
scalar function $f$
\begin{equation} \label{eq:enrichment}
  \bm{u} =
  \underbrace{
    \bm{\overline{u}}  + \bm{\left[ u \right]}f
  }_{\bm u_c} +
    \underbrace{
      \bm{\left[ u \right]} \left( H_\Gamma - f \right)
  }_{\bm u_{nc}}
\end{equation}
so that
$\bm u_{c}$ is the conformal part of displacement,
$\bm u_{nc}$ is the non-conformal part of displacement \cite{simo1994new}
as shown in Fig.~\ref{fig:enrichment}c.
The form of the level-set (or ramp) function $f$ depends on the discretization
and is discussed elsewhere \cite{oliver2003study,foster2007embedded}.
An expression for the small strain $\bm{\varepsilon}$ can be obtained by taking
a symmetric gradient of Eq \ref{eq:enrichment}:
\begin{equation} \label{eq:strain-full}
  \bm{\varepsilon} = \nabla^s \bm{u} =
  \nabla^s \bm{u}_c -
  \left(  \bm{\left[ u \right]} \otimes \nabla f \right)^s +
  \left( H_\Gamma - f \right) \nabla^s\bm{\left[ u \right]} +
  \left(  \bm{\left[ u \right]} \otimes \bm{n} \right)^s \delta_\Gamma
\end{equation}
where $\delta_\Gamma$  is the  Dirac delta-function and  $\bm{n}$ is  the unit
normal  vector to  the fracture  surface. We  neglect the  second term  in the
right-hand  side of  Eq.  \ref{eq:strain-full} due  to the  jump-discontinuity
assumption, so that Eq. \ref{eq:strain-full} becomes \cite{mosler2005novel}
\begin{equation} \label{eq:strain}
  \bm{\varepsilon} =
  \nabla^s \bm{u}_c -
  \left(  \bm{\left[ u \right]} \otimes \nabla f \right)^s +
  \left(  \bm{\left[ u \right]} \otimes \bm{n} \right)^s \delta_\Gamma
\end{equation}

\paragraph{Plasticity formulation for SDA:}
%
% Following a number of authors,
We incorporate SDA into
the non-associated plasticity framework described by the system
\cite{oliver1996modelling,mosler2005novel,foster2007embedded}:
\begin{subequations} \label{eq:regular-plasticity}
\begin{align}
  & \dot{\bm{\sigma}} = \varmathbb{C} : \left( \dot{\bm{\varepsilon}} - \dot{\bm{\varepsilon}}^p \right) \\
  & \dot{\bm{\varepsilon}}^p = \lambda  \frac{\partial G}{\partial \bm{ \sigma }} \\
  & \dot{q} = \lambda H \frac{\partial G}{\partial q} \\
  & F\left(\bm{\sigma}, q \right) = 0
\end{align}
\end{subequations}
where $\dot{\bm{\varepsilon}}^p$ is the plastic strain,
$\lambda$ is the plastic multiplier,
$q$ is the stress-like internal variable,
$G$ and $F$ are the plastic potential and yield function, respectively,
and $H$ is the softerning/hardening parameter.
In system \eqref{eq:regular-plasticity},
Eq. \ref{eq:regular-plasticity}a is the stress-strain relation.
Eq. \ref{eq:regular-plasticity}b described the plastic strain evolution.
Eq. \ref{eq:regular-plasticity}c is the softening/hardening law, and
Eq. \ref{eq:regular-plasticity}d is the yield surface.
We emphasize that the system \eqref{eq:regular-plasticity}
% is given in terms of the effective stresses $\bm\sigma$.
contains the effective stress $\bm\sigma$ as opposed to the total
stress $\bm{S}$ \cite{callari2010strong}.

The governing system of equation for SDA can be obtained by substituting
Eq.~\ref{eq:strain} into system \eqref{eq:regular-plasticity},
assuming
that the stress $\bm{ \sigma }$ is bounded
and that the distributions of the plastic multiplier
$\lambda = \lambda_\delta \delta_\Gamma$ and hardening modulus $H = H_\Gamma \delta_\Gamma$
are singular
% the stress boundness \cite{foster2007embedded} and
% the singularity of distribution of the plastic multiplier
% $\lambda = \lambda_\delta \delta_\Gamma$ and hardening modulus $H_\Gamma = H \delta_\Gamma$
\cite{simo1993analysis,mosler20033d}:
\begin{subequations} \label{eq:sda-system}
  \begin{align}
    & \dot{\bm{\sigma}} = \varmathbb{C} : \left[ \nabla^s \dot{\bm{u_c}} -
          \left( [\dot{\bm{u}}] \otimes \nabla f \right)^s \right]  \\
    & \left[\dot{\bm{u}}\right] = \lambda_\delta \frac{\partial G}{\partial \bm{t}} \\
    & \dot{q} = -\lambda_\delta H_\delta \frac{\partial G}{\partial q} \\
    & F\left( \bm{t}, q \right) = 0
  \end{align}
\end{subequations}

\subsubsection{Forms of plastic potential} \label{sec:potentials}
In  this section,  we specify  the  flow rule  and plastic  potential for  three
possible states of a fracture: slip, opening,  and stick.

\paragraph{Slip:}
During slip, the jump vector in the fracture is nearly-parallel to its
surface.
We use the following flow rule and plastic potential to impose friction in the fracture:
% We use the following frictional forms for the flow rule and plastic potential:
%
\begin{subequations} \label{eq:shear-flow-and-potential}
\begin{align}
    F \left(\bm{t}, q \right) = t_\tau - \mu t_n - q \\
    G \left(\bm{t}, q \right) = t_\tau - \theta t_n - q
 \end{align}
\end{subequations}
where
$t_\tau = \bm{n}\cdot\bm{\sigma}\cdot \bm{\tau}$ is the tangent traction,
$t_n = \bm{n}\cdot\bm{\sigma}\cdot \bm{n}$ is the normal traction,
$\mu$ is the tangent of the friction angle,
and $\theta$ is the tangent of the dilation angle.
Eq. \ref{eq:shear-flow-and-potential}b transforms
Eq. \ref{eq:sda-system}b into
\begin{equation} \label{eq:sda-shear-derivative}
  \begin{aligned}
    & \left[\dot{u}\right]_n = \theta \lambda \\
    & \left[\dot{u}\right]_\tau =  \lambda
  \end{aligned}
\end{equation}
where $\left[\dot{u}\right]_n$ and $\left[\dot{u}\right]_\tau$
are the normal and tangential components of $\left[\dot{\bm{ u }}\right]$.

\paragraph{Opening:}
For an open fracture, we propose to use three components of the jump function
$\left[u\right]$ as independent local variables.
We select the flow rule and plastic potential to
enforce
the absence of all traction components independently:
\begin{equation} \label{eq:sda-opening}
  \begin{aligned}
    F_1(\bm{t}, q) = G_1(\bm{t}, q) = t_n - q \\
    F_2(\bm{t}, q) = G_2(\bm{t}, q) = t_{\tau_1} - q \\
    F_3(\bm{t}, q) = G_3(\bm{t}, q) = t_{\tau_2} - q \\
  \end{aligned}
\end{equation}
Note that in the case of an opening fracture the plastic potential $G$ and
flow rule $F$ are equal (associative plasticity) and have three components.
This choice of the plastic potential transforms Eq.~\ref{eq:sda-system}b
to the following form:
\begin{equation} \label{eq:opening-derivative}
  \begin{aligned}
    \left[\dot{u}\right]_{\tau_1} = \lambda_1 \\
    \left[\dot{u}\right]_{\tau_2} = \lambda_2 \\
    \left[\dot{u}\right]_n = \lambda_3 \\
  \end{aligned}
\end{equation}

\paragraph{Stick:}
The  case of  a closed  non-slipping fracture  (stick case)
also requires  special treatment.
Although during loading, a rock with a fracture behaves elastically
until the yield  conditions are met, it is important to
enforce the zero-normal jump and no-slip  conditions during the unloading of a
fracture:
\begin{equation} \label{eq:sda-stick}
  \begin{aligned}
    \left[ u \right]_n = 0 \\
    \left[\dot{u}\right]_{\tau} = 0 \\
  \end{aligned}
\end{equation}
We emphasize that we set  the normal jump and tangential jump rate to zero.

\section{Discretization}
Our numerical model for embedded  fractures utilizes a hybrid formulation with
first-order  Galerkin  displacement  approximation,  Petrov-Galerkin  for  the
enhanced   strain   approximations,   and   piecewise-constant   finite-volume
representation of pressure.  The distribution of variables in a  cell is shown
in Fig.~\ref{fig:edfm-discretization}.  Due to the choice  of shape functions,
the  conforming  displacement  is  defined  at grid  nodes.  As  discussed  in
Section~\ref{sec:mechanics}, the displacement jump  $\bm{\left[ u \right]}$ on
the  fracture  is  assumed  to  be discontinuous  and  is  approximated  as  a
piecewise-constant  function. We  define  the displacement  jump  at the  cell
centers,  although, as  discussed in  Section~\ref{sec:sda-linearization}, its
location is  irrelevant. Fluid  pressure in  the rock  matrix and  fracture is
defined in the matrix and fracture control volume centers, respectively.

\begin{figure}[ht]
  \centering
  \includegraphics[width=0.2\textwidth]{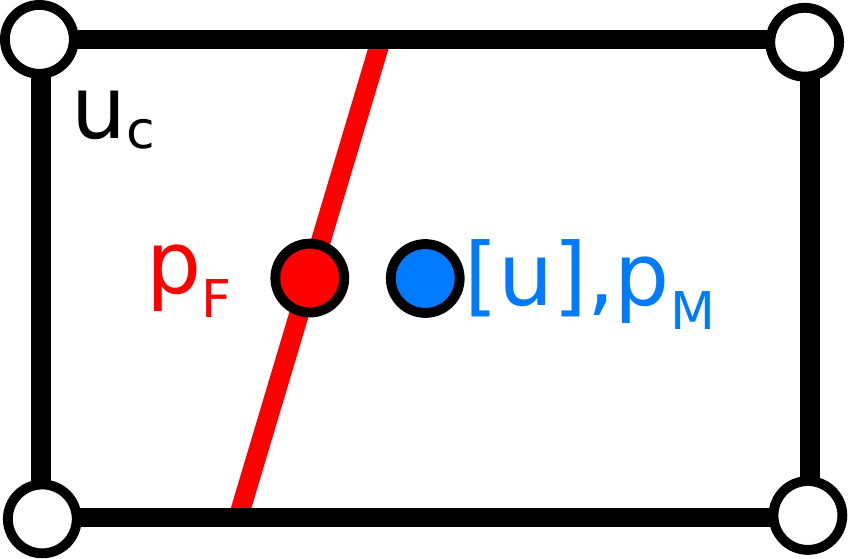}
  \caption{\label{fig:discretization}
    Location of variables in the coupled formulation.
    The conforming displacement $\bm u_c$ is defined in cell vertices.
    The jump $\bm{\left[ u \right]}$ and the fluid pressure in the rock matrix $p_M$ are defined at the cell centers.
    The fluid pressure $p_M$ in the fracture is defined at the center of the fracture control volume.
  }
\end{figure}

In the following sections, we provide  the relevant details regarding the  numerical treatment of
fractures.  We first  discuss  fluid  flow and  then  provide  the details  of
discretizing the mechanics equilibrium equations.

\subsection{Fluid flow in fractures} \label{sec:frac-flow}
The EDFM approach for approximating  fluid flow flow consists of approximating
the matrix-fracture flow $q_{MF}$ and fracture-fracture flow $q_{FF}$ terms in
Eq.~\ref{eq:matrix-flow}  and  \ref{eq:fracture-flow}.  We  opt  to  select  a
simplified approach first introduced  in \cite{li2008efficient}. More accurate
schemes are available as discussed in Section \ref{flow-limitations}.

\begin{figure}[ht]
  \centering
  \includegraphics[width=0.5\textwidth]{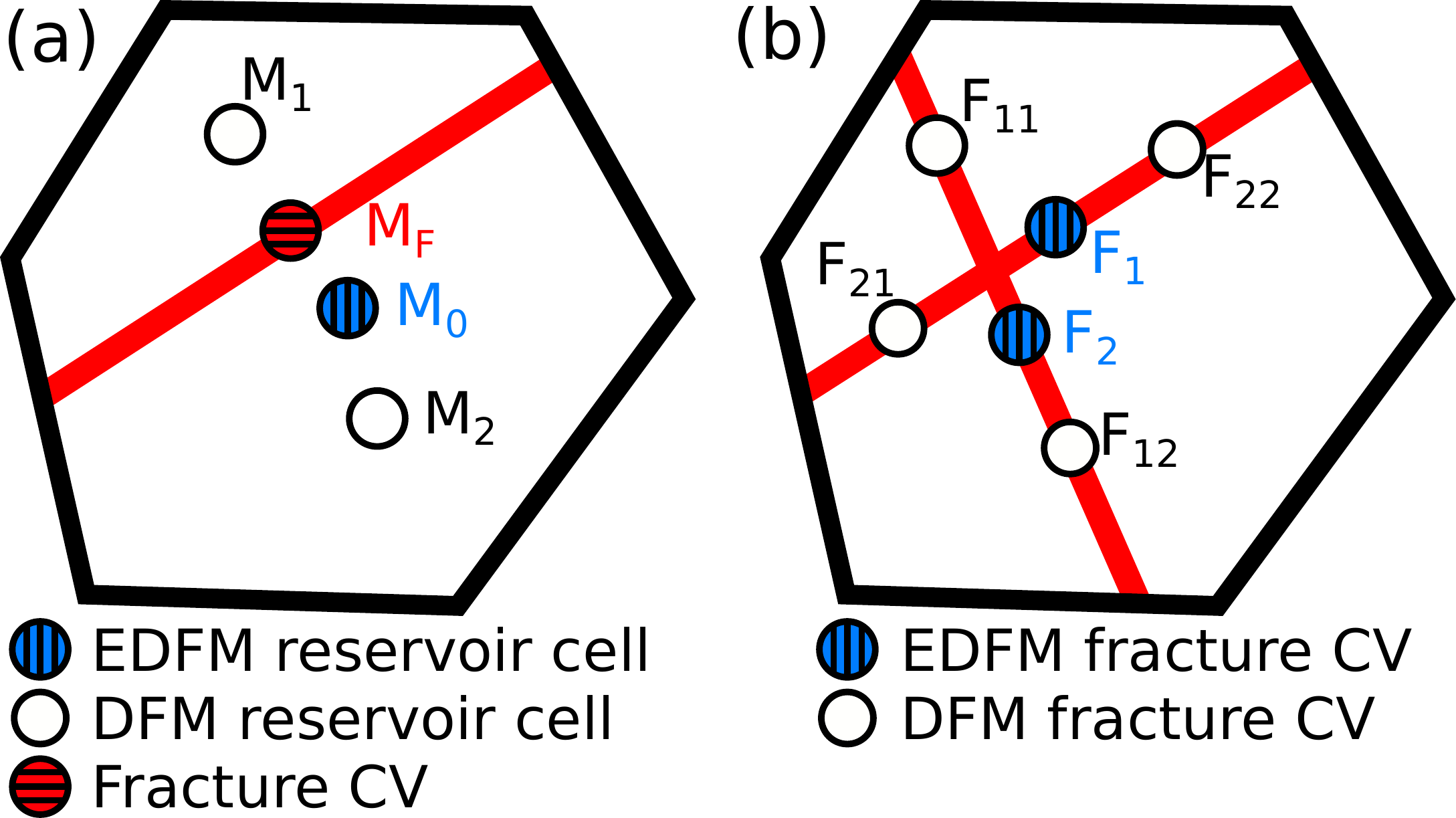}
  \caption{\label{fig:edfm-discretization}
    Fluid domain discretization.
    (a) Control volumes (CV) for EDFM and DFM models.
    (b) Schematics for fracture-fracture transmissibility calculation.
  }
\end{figure}

We  compute the  total mass  flux  between a  reservoir control  volume and  a
fracture control volume as:
\begin{equation} \label{eq:MF-flow}
  Q_{MF} = \frac{2A \ \bm{n} \cdot \bm{k_R}\cdot\bm{n}}{\overline{d}}
  \frac{\rho_f}{\mu_f}
  \left( p_M - p_F \right),
\end{equation}
where the coefficient $\overline{d}$ is the average distance between the points
in the reservoir cell and the fracture cells.
When a fracture element with its center  in $\bm{M_F}$ bisects a cell with the
volume $V_0$ and the center in  $\bm{M_0}$, then the initial volume splits into
shape~I with the  center in $\bm{M_1}$ and volume $V_1$  and shape~II with the
center       in       $\bm{M_2}$       and       volume       $V_2$       (see
Fig.~\ref{fig:edfm-discretization}). Obviously, the sum  of the volumes of the
shapes I and II  is equal to the original cell volume $V_1  + V_2 = V_0$. Then
the coefficient $\overline{d}$ can be evaluated as
\begin{equation} \label{eq:averaging}
  \overline{d} = \frac{V_1}{V_0} || \bm{M_1} - \bm{M_F} || +
                 \frac{V_2}{V_0} || \bm{M_2} - \bm{M_F} ||
\end{equation}

We  next  discuss  the flux  term  that  arises  at  the intersection  of  two
fractures. We assume that the  flux between two intersecting fracture elements
$F_1$   and  $F_2$   is   proportional  to   the   pressure  difference   (see
Fig.\ref{fig:edfm-discretization}b):
\begin{equation} \label{eq:FF-flow}
 Q_{12} = T^1_2  \frac{\rho_f}{\mu} \left( p_1 - p_2  \right)
\end{equation}
Hereafter, we denote the transmissibility between  two segments $i$ and $j$ by
$T^i_j$.  To compute  the  transmissibility $T^1_2$  between two  intersecting
fracture elements $F_1$ and $F_2$, we split these elements by the intersection
line  into  the  segments  $F_{11}$   ,  $F_{12}$  and  $F_{21}$  ,  $F_{22}$,
respectively  Fig.~\ref{fig:edfm-discretization}b).  Then   the  transmissibility
$T^1_2$ is obtained with sum of the corresponding transmissibilities:
\begin{equation} \label{eq:FF-trans}
  T^1_2 = T^{11}_{22} + T^{11}_{21} + T^{12}_{22} + T^{12}_{21},
\end{equation}
where the transmissibilities between the fracture segments (in the right-hand
side of Eq. \ref{eq:FF-trans}) are computed with the star-delta transformation
\cite{karimi2003efficient}.

\subsection{Mechanics} \label{sec:sda-linearization}

% \subsubsection{Level-set function}
% \timur[We use the shape functions only once. Could we define the function 'f' earlier with the relevant citation?]
% The function $f$ introduced in Eq.~\ref{eq:enrichment} is a continuous
% approximation of the Heaviside step function $H_\Gamma$.
% According to multiple authors, a convenient form to represent $f$ is
% \begin{equation} \label{eq:level-set}
%   f \left( \bm{x} \right) = \sum_{i=1}^{N^e_n} N_a \left( \bm{x} \right)
%   H_\Gamma \left( \bm{x} \right)
% \end{equation}
% where $N_a   \left( \bm{x} \right)$ are the first-order
% Continuous Galerkin shape functions, and $N^e_n$ is the number of nodes in
% an element \cite{oliver2003study,foster2007embedded}.

\subsubsection{Stress continuity}
We now  discuss the discretization of  the mechanical system that  consists of
Eq.~\ref{eq:cauchy}   and    the   system    \eqref{eq:sda-system}.
Using the enhanced assumed strain approach, we apply finite-element approximations to Eq.
\ref{eq:cauchy} \cite{simo1990class}:
\begin{subequations} \label{eq:lin-cauchy}
  \begin{align}
    \int_{\Omega^e} \nabla^s \bm{\eta} : \bm{S}\ d\Omega  &= \int_{\Omega^e}\rho \bm{g} d\Omega, \\
    \int_{\Omega^e} \bm{\gamma} : \bm{S}\ d\Omega &= 0,
  \end{align}
\end{subequations}
where $\bm{\eta}$ denotes a continuous test function,
$\bm{\gamma}$ is the variation of the enhanced strain,
and $\Omega^e$ is an element that contains a fracture.
Eq. \ref{eq:cauchy} transforms into a system of two equations due to singular strain
across the fracture.
% Eq.~\ref{eq:lin-cauchy}b is the result of the singularity in the strain field.
% Eq. \ref{eq:cauchy} results
%
To solve  the coupled system \eqref{eq:cauchy}  and \eqref{eq:matrix-flow}, we
use  the  contact  integral   in  Eq.~\ref{eq:lin-cauchy}b  as  an  additional
constraint.  This  expression must  be  modified  for  the case  of  piecewise
constant jump values  \cite{borja2000finite} and can be replaced  by the local
stress-continuity condition \cite{regueiro1999finite,callari2010strong}:
\begin{equation} \label{eq:stress-avg}
  \displaystyle{\frac{1}{V^e}}\int_{\Omega^e} \bm{S} \cdot \bm{n}\ d\Omega = \bm{t} + p_f \bm{n}
\end{equation}
Eq. \ref{eq:stress-avg} relates the average total stress $\overline{\bm{S}}$
in an element to the fracture traction $\bm{ t }$ and fracture pressure $p_F$.
Hereafter, we use the following notation
$\overline{(\bullet)} := \displaystyle{\frac{1}{V^e}}\int_{\Omega^e} (\bullet) d\Omega$.
Therefore, in order to complete the system \eqref{eq:sda-system},
we use the volume average of Eq. \ref{eq:sda-system}a
\begin{equation} \label{eq:support-point-stress}
 \dot{\overline{\bm{\sigma}}} = \varmathbb{C} :
\left[\nabla^s \dot{\bm{u}}_{\bm c} -
  \left( \left[ \dot{\bm{u}} \right] \otimes \overline{\nabla f}  \right)^s
\right],
\end{equation}
and  the  following relationship  between  the  average effective  stress  and
fracture traction:
\begin{equation} \label{eq:stress-traction}
  \bm{t} + p_F \bm{n} = \bm{\overline{\sigma}} \cdot n + b p_M \bm{n}.
\end{equation}
We emphasize  that the  average effective stress  $\overline{\bm{\sigma}}$ and
jump $\bm \left[ u \right]$ are functions of the matrix and fracture pressures
and  have the  corresponding terms  in linearization  as discussed  in Section
\ref{sec:linearization}.  Note the  absence  of the  overline  above the  jump
$\bm{\left[ u  \right]}$ as $\bm{\left[  u \right]} =  \bm{\left[ \overline{u}
\right]}$ due to the piecewise-constant jump assumption.

In our  formulation, we solve  the local system of  equations \eqref{eq:sda-system},
\eqref{eq:support-point-stress},  and  \eqref{eq:stress-traction} for
all the elements containing fractures using the Newton-Raphson method.
We  use  slip  rate  $\bm{\left[  \dot{u}  \right]}$,  average  stress  tensor
$\overline{\bm{\sigma}}$,  and the  Lagrange multiplier  $\lambda$ as  primary
unknowns. Nodal displacement $\bm u_c$, matrix and fracture pressure $p_M$ and
$p_F$  are fixed  during the nonlinear  iterations. In  order to  obtain quadratic
convergence  for   the  system  \eqref{eq:lin-cauchy},  a   consistent  stress
linearization must be performed \cite{borja2013plasticity}.

\subsubsection{Stress Linearization} \label{sec:linearization}
Eq. \ref{eq:support-point-stress} and Eq. \ref{eq:sda-system}b-d constitute a complete
system, which is used to find the jump $\left[ \bm{u} \right]$ and traction vector $\bm{t}$
from the strain $\varepsilon$.
After computing these quantities, the effective stress $\sigma$  in Gauss quadrature points  can be
obtained with  Eq.~\ref{eq:sda-system}a.  The effective-stress derivatives  can be
computed from the following linearization:
\begin{equation} \label{eq:sda-linearization}
  \text{d}\bm{\sigma} = \varmathbb{C} :
  \left[
    \textbf{I} +
  \left(
    \frac{\partial \bm{\left[ u \right]}}{ \partial \overline{\bm\varepsilon}} \otimes \nabla f
  \right)^s
  \right]
   \text{d} \bm{\varepsilon}
  +
  \varmathbb{C} : \left(
    \frac{\partial \bm{\left[ u \right]}}{ \partial p_f} \otimes \nabla f
  \right)^s \text{d} p_f
  +
  \varmathbb{C} : \left(
    \frac{\partial \bm{\left[ u \right]}}{ \partial p_m} \otimes \nabla f
  \right)^s \text{d} p_m .
\end{equation}
The derivatives
$\frac{\partial \bm{\left[ u \right]}}{ \partial p_f}$ and
$\frac{\partial \bm{\left[ u \right]}}{ \partial p_m}$ can be obtained from the system
\eqref{eq:sda-system}.
To evaluate these derivatives and obtain quadratic convergence of the Newton scheme,
we use an approach based on the inverse theorem
and  automatic differentiation as discussed elsewhere \cite{garipov2018unified}.
Using Eq.~\ref{eq:effective}, we  write the  linearization for the  total stress
tensor:
\begin{equation} \label{eq:linearization-total-stress}
  \text{d}\bm{S} =  \varmathbb{C} :
  \left[
    \textbf{I} +
  \left(
    \frac{\partial \bm{\left[ u \right]}}{ \partial \overline{\bm\varepsilon}} \otimes \nabla f
  \right)^s
  \right]
   \text{d} \bm{\varepsilon}
  +
  \varmathbb{C} : \left(
    \frac{\partial \bm{\left[ u \right]}}{ \partial p_f} \otimes \nabla f
  \right)^s \text{d} p_f
  +
  \varmathbb{C} :
  \left[
  \left(\frac{\partial \bm{\left[ u \right]}}{ \partial p_m} \otimes \nabla f \right)^s
  + b \textbf{I}
  \right]
  \text{d} p_m
\end{equation}

We emphasize that the linearization of the total stress with respect to the matrix pressure $p_m$
is not isotropic due to the term
$\left(\frac{\partial \bm{\left[ u \right]}}{ \partial p_m} \otimes \nabla f\right)^s$.
%
% To compute the additional derivatives we apply the inverse theorem approach
% \cite{garipov2018unified}.
% A consistent linearization of  the governing equations allows us to achieve
% quadratic convergence in a damped Newton scheme \cite{mosler2005novel}.

\section{Results} \label{sec:results}
In this section  we validate presented mechanical  model with plane-strain %
analytical solutions and compare its spatial  convergence with that of the DFM
model.
The  simulation cases presented  in this section  are computed
in a domain with a single cell in the third (z) direction since
% on  a 2D square domain (one layer of cells) as
the available  analytical solutions are
derived based on the plain-strain assumption.  The domain size was assumed ten
times larger than  the fracture length to mimic an  infinite plane. Further we
use the  term EDFM  (embedded discrete  fracture model) to  refer to  both the
mechanical and  flow models. In the  context of mechanics, the  terms EDFM and
SDA are interchangeable.

\subsection{Mechanical tests}
%
% First, we  consider a  shear slipping  test problem  with an  inclined fracture
First, we  consider an  inclined fracture slipping
under compressive loading conditions. Here, a single fracture with length $l$ = 10 m
and striking at $\alpha$ = 30$\degree$ is slipping due to the applied compressive
load $\sigma_x$ = 10 MPa ($\sigma_y$ = 0). The tangential slip on the
fracture has a parabolic profile \cite{phan2003symmetric}:
\begin{equation} \label{eq:analytical-shear}
  \left[ u_{\tau} \right] = \frac{4\left( 1-\nu^2 \right)}{E}
  \sigma_x \sin \alpha \left( \cos \alpha - \mu \sin \alpha \right)
  \sqrt{l^2 - \left( x - l \right)^2}
\end{equation}
where $E$ is the rock Young's modulus and $\nu$ is the rock Poisson's ratio.

Second, we consider an open fracture test. In this case, an x-aligned fracture
($\alpha=0$)  opens due  to  the  applied tensile  load  $\sigma_y$  = 10  MPa
($\sigma_x$ = 0). This problem is analogous to the well-know Sneddon's problem
of   a    pressurized   fracture    in   a   plane-strain    infinite   domain
\cite{sneddon1946opening}. The  analytical solution for the  fracture aperture
as a function of the x-coordinate is given by:
\begin{equation} \label{eq:analytical-sneddon}
  \left[ u_n \right] = \frac{1-\nu}{\mu} \sigma_y \sqrt{l^2 - \left( x - l \right)^2}
\end{equation}
where $\mu$ is the rock shear modulus.
All the geometrical and mechanical parameters used in the simulations
are listed in Table \ref{tab:parameters-mech-simulation}.

\begin{figure}[H]
    \centering
    \includegraphics[width=0.7\textwidth]{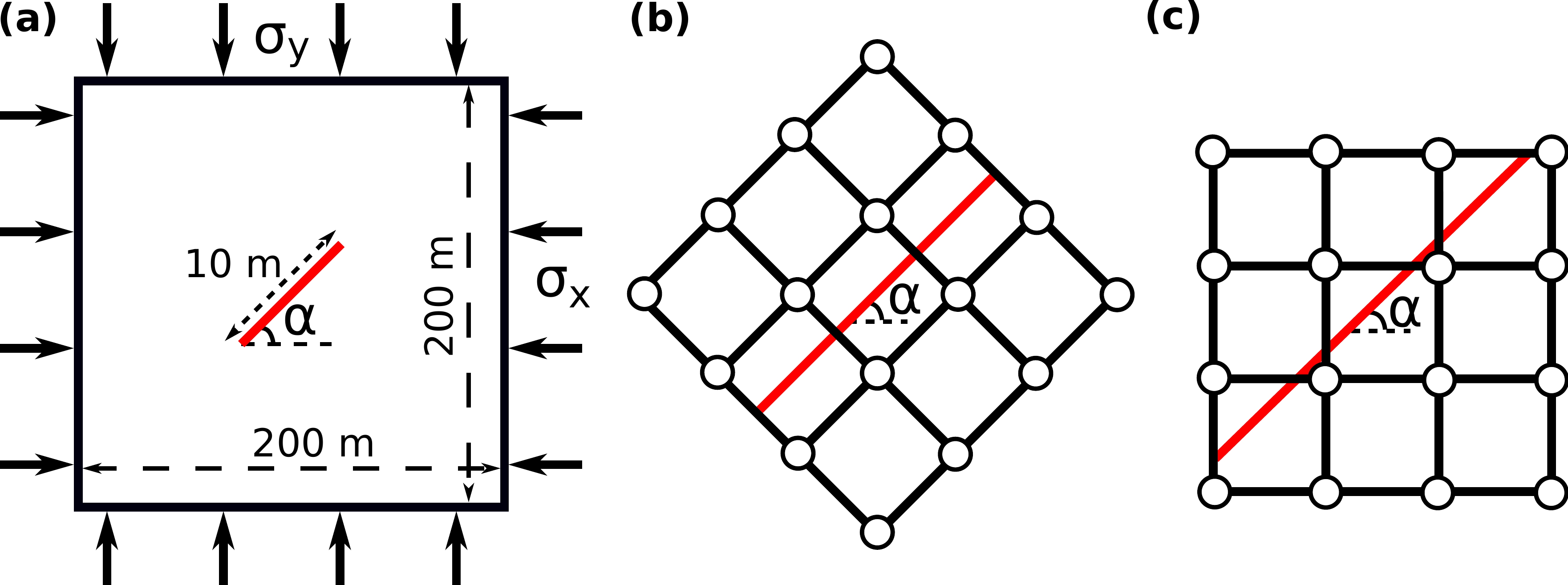}
    \caption{
      (a) Simulation domain.
      (b) A fracture conforming to the grid.
      (c) A fracture not conforming to the grid.
	}
	\label{fig:domain-test-1frac}
\end{figure}

\begin{table}[H]
  \centering
  \begin{tabular}{lc}
    \hline
    Property                              & Value                        \\
    \hline
    Domain size [m$^3$]                   & 200 $\times$ 200 $\times$ 10 \\
    Fracture length [m]                   & 10                           \\
    Matrix Young's modulus $E$ [MPa]      & 10$^3$                       \\
    Matrix Poisson's ratio $\nu$ [-]      & 0.25                         \\
    Friction coefficient $\mu$ [-]        & 0.6                          \\
    Dilation coefficient $\theta$ [-]     & 0.0                          \\
    Initial cohesive strength $q_0$ [MPa] & 0.0                          \\
    Hardening parameter $H$ [MPa]         & 0.0                          \\
    \hline
  \end{tabular}
  \caption{
    Geometrical and material parameters used in the mechanical simulations.
  }
  \label{tab:parameters-mech-simulation}
\end{table}

\subsubsection{Grid conforming to the fracture} \label{sec:conformal}
\begin{figure}[H]
    \centering
    \includegraphics[width=\textwidth]{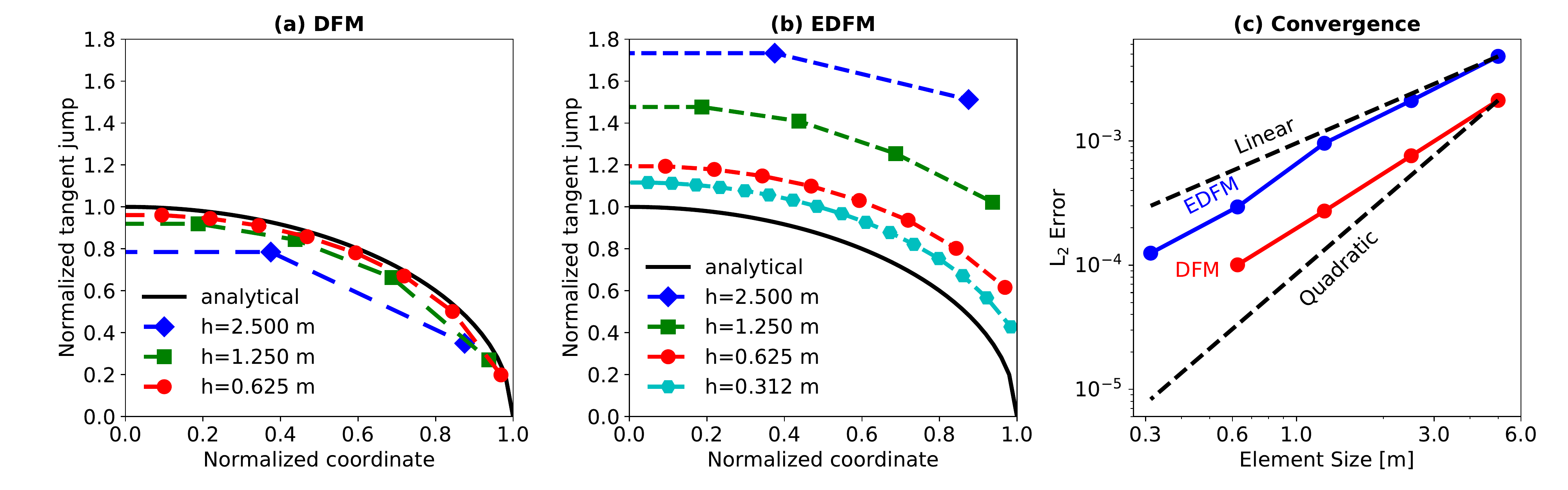}
    \caption{
      Spatial convergence exhibited by the EDFM and DFM models for the shear problem.
      (a) Tangential jump as a function of coordinate obtained at various refinement levels.
      Note that DFM model underestimates the jump, whereas EDFM overestimates it.
      (b) L$_2$-error in the tangential jump value as a function of the element size h.
      Both models manifest super-linear convergence. The error in the tangential jump obtained with
      the DFM model is lower than that obtained with the EDFM model.
	}
	\label{fig:shear-convergence}
\end{figure}

We  first present  modeling results  obtained on a domain with the grid
conforming to the fracture (Fig.  \ref{fig:domain-test-1frac}b).  We  performed
several simulations  on hexahedral grids
% with various  element sizes: h  = 5, 2.5, 1.25, 1.1875, and 0.3125 m.
with various numbers of cell elements containing the fracture $n_F$ = 4, 8, 16, and 32.
Identical cases with  the same grid element sizes were  simulated with the DFM
model. The  tangential jump  on the  fracture surfaces given  by EDFM  and DFM
models at various grid refinement levels,  as well as the analytical solution,
are  shown  in  Fig.   \ref{fig:shear-convergence}a-b.  All  slip  values  are
normalized by  the maximum  of Eq.~\ref{eq:analytical-shear}.  The coordinates
are normalized by the fracture length $l$. The solution obtained from the EDFM
model  overestimates  the  jump  value  and  converges  monotonically  to  the
analytical  solution  (Fig.~\ref{fig:shear-convergence}a).  In  contrast,  the
solution  obtained  from  the  DFM  model underestimates  the  jump  and  also
converges       monotonically      to       the      analytical       solution
(Fig.~\ref{fig:shear-convergence}b).

The  L$_2$-error in  the solutions  produced  by the  EDFM and  DFM models  as
functions     of     the     grid     element     size     is     shown     in
Fig.~\ref{fig:shear-convergence}b. Both  errors given  by DFM and  EDFM models
reduce while  refining the grid  at approximately the  same rate. In  the same
figure, we  provided linear and  quadratic trends  to illustrate the  order of
spatial convergence  of the models.  The errors given  by DFM and  EDFM models
exhibit super-linear convergence.

\begin{figure}[H]
    \centering
    \includegraphics[width=\textwidth]{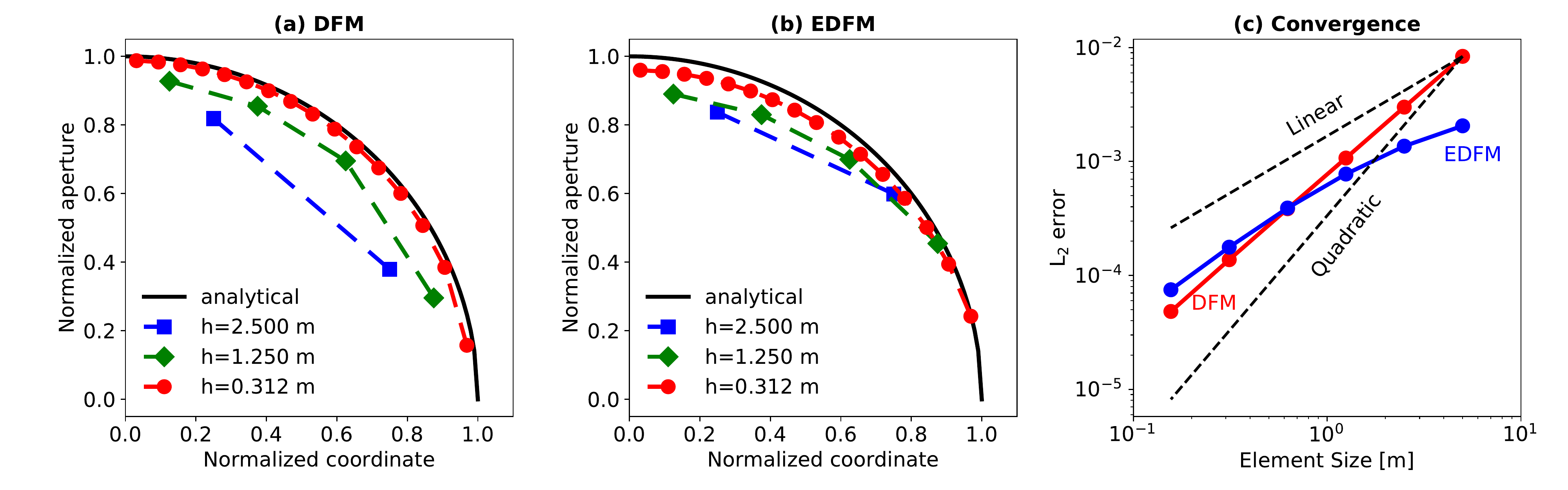}
    \caption{
      Spatial convergence manifested by the EDFM and DFM models for the opening problem.
      (a-b) Fracture aperture as a function of the coordinate obtained with
      the DFM (a) and EDFM (b) models.
      (c) L$_2$-error in the fracture aperture as a function of the grid element size.
      DFM model manifests super-linear convergence.
      EDFM model exhibits convergence that is worse than linear trend on
      coarse grids and linear convergence on fine grids.
	}
	\label{fig:sneddon-convergence}
\end{figure}

For the second test problem, we  also performed a convergence analysis of the
DFM and EDFM  models on Cartesian grids
% with element sizes h =  5, 2.5, 1.25, 1.1875, and 0.3125 m.
with the numbers of grid comprising the fracture $n_F$ = 2, 4, 8, 16, and 32.
The fracture aperture given by the  EDFM and DFM models
at various  grid refinement levels,  as well  as the analytical  solution, are
shown in Fig.~\ref{fig:sneddon-convergence}a-b.  Both models underestimate the
aperture value and converge monotonically to the analytical solution.

The L$_2$-errors  in fracture aperture as  functions of the grid  element size
are  shown in  Fig. \ref{fig:sneddon-convergence}c.  Same as  in the  previous
case,  the  results   obtained  from  the  DFM   model  manifest  super-linear
convergence. In contrast,  the EDFM convergence in this case  is different: at
coarse grids the L$_2$ error $e$ as a function the of element size $h$ has the
slope  less steep  than the  linear  trend; at  fine  grids the  slope of  the
convergence  curve is  super-linear  ($e \propto  h^{1.24}$). Another  notable
difference with  the case of  a sliding fracture  is that the  numerical error
given  by  the EDFM  model  at  coarse grids  (less  than  eight elements  per
fracture) is less than that of the  DFM model. At fined grids (more than eight
elements  per fracture)  the DFM  model  predicts the  fracture aperture  more
accurately than the EDFM model.

\subsubsection{Grid not conforming to the fracture}

In this section,  we present EDFM simulation results for  the cases of slipping
and opening  fractures with  the problem  setup identical  to that  in Section
\ref{sec:conformal}.  The   only  difference  with   the  previously-presented
simulations  is  the  grid,  which  does not  conform  to  the  fracture  (see
Fig.~\ref{fig:domain-test-1frac}c).

\begin{figure}[H]
    \centering
    \includegraphics[width=\textwidth]{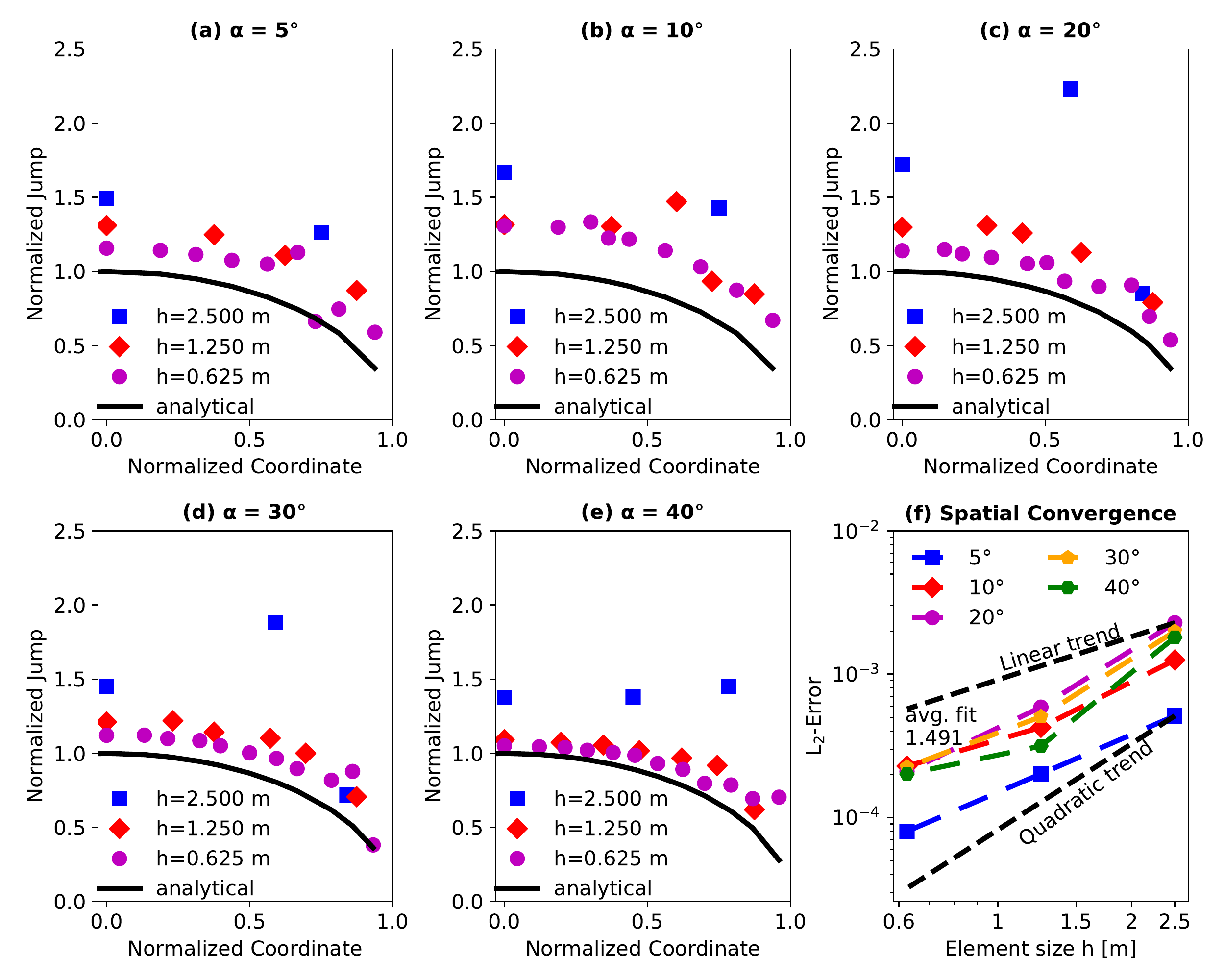}
    \caption{
      Spatial convergence of the EDFM mode on non-conforming mesh for the fracture slip problem.
      (a-e) Normalized fracture slip as a function of normalized distance along fracture
      at various refinement level for various fracture strike angles $\alpha$.
      (f) L$_2$-error in tangential jump as a function of mesh element size
      for various fracture strike angles $\alpha$.
	}
	\label{fig:edfm-nc-shear}
\end{figure}

For the  case of  a slipping  fracture we  performed numerical  simulations on
Cartesian grids  with the element sizes  $h$ = 2. m,  1.25 m, and 0.625  m. In
these  simulations,  we also  varied  the  fracture  strike angle  $\alpha$  =
5$\degree$, 10$\degree$, 20$\degree$, 30$\degree$, and 40$\degree$ in order to
eliminate the effect  of fracture orientation. The spatial  convergence of the
EDFM  model for  the case  of a  slipping fracture  on non-conforming  mesh is
summarized  in Fig.  \ref{fig:edfm-nc-shear}. Fig.  \ref{fig:edfm-nc-shear}a-e
show  the slip  along the  fracture normalized  by the  maximum slip  given by
Eq.~\ref{eq:analytical-shear} as  a function of a  normalized coordinate along
the fracture for particular fracture orientations.
Fig. \ref{fig:edfm-nc-shear}a-e indicate that the  fracture slip given by EDFM
on fine  grids ($h=$0.625 m)  is closer to  the analytical solution  than that
obtained on  the coarse grids ($h=$2.5  m) for each fracture  strike $\alpha$.
Fig. \ref{fig:edfm-nc-shear}f shows the  L$_2$-error in the numerical solution
as a function of the grid element  size at various fracture strike angles. The
error in  the numerical  solution decreases  monotonically while  refining the
mesh for each fracture orientation. The  average polynomial fit of the data in
Fig.~\ref{fig:edfm-nc-shear}f  yields $e  \propto  h^{1.49}$, which  indicates
super-linear convergence.

\begin{figure}[H]
    \centering
    \includegraphics[width=\textwidth]{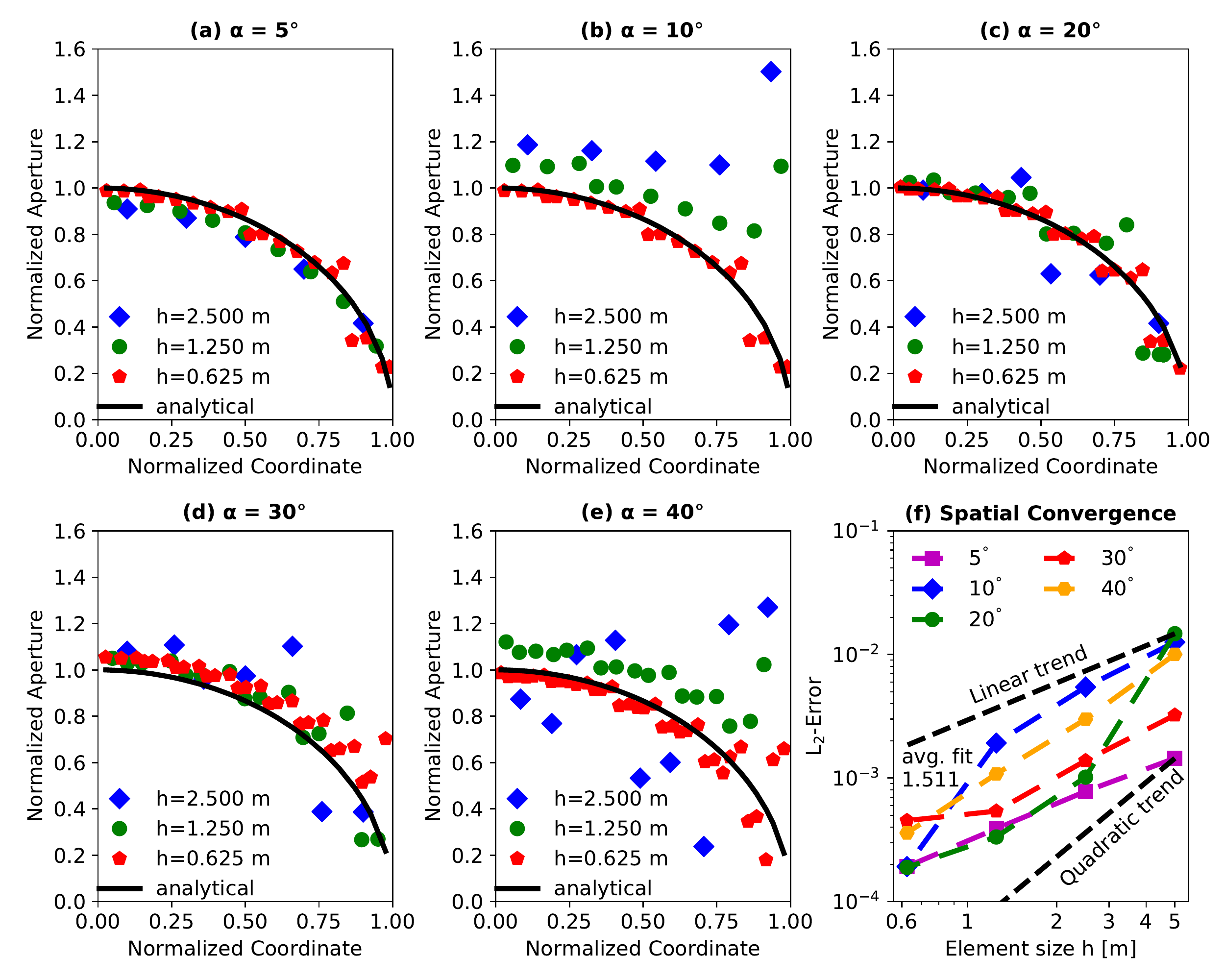}
    \caption{
      Spatial convergence of the EDFM mode on non-conforming mesh for the fracture opening problem.
      (a) Normalized fracture aperture as a function of normalized distance along fracture.
      (b) L$_2$-error in fracture aperture as a function of mesh element size.
	}
	\label{fig:edfm-nc-sneddon}
\end{figure}

We now present the  results of a convergence study for the  case of an opening
fracture on  non-conforming grids. Fig. \ref{fig:edfm-nc-sneddon}a-e  show the
normalized fracture  aperture as a  function of the normalized  coordinate for
various  grid element  sizes  and orientations.  Similarly to  the  case of  a
slipping fracture,  the EDFM model captures  poorly the shape of  the aperture
profile on coarse  grids (approximately 3-5 elements per  fracture). Upon grid
refinement, however, the numerical results converge to the analytical solution
for each case of grid orientation.

The spatial  convergence of  the EDFM  model on  non-conforming grids  for the
Sneddon's  problem is  shown in  Fig.~\ref{fig:edfm-nc-sneddon}f.  The spatial
convergence  is monotonic,  nonlinear,  and differs  for various  orientation
cases. On average, however, the error decreases super-linearly with a decrease
in the mesh element size as estimated from the polynomial fit of the data in
Fig.~\ref{fig:edfm-nc-sneddon}f $e \propto h^{1.51}$.

Overall, the proposed  EDFM mechanical model is accurate and  converges to the
analytical  solution.  We observe  super-linear  convergence  behavior in  all
simulation cases  on both conforming  and non-conforming grids. The  DFM model
also converges  super-linearly and  is more  accurate than  the EDFM  model on
conforming grids.

\subsection{Fluid flow test}
\begin{figure}[H]
    \centering
    \includegraphics[width=0.3\textwidth]{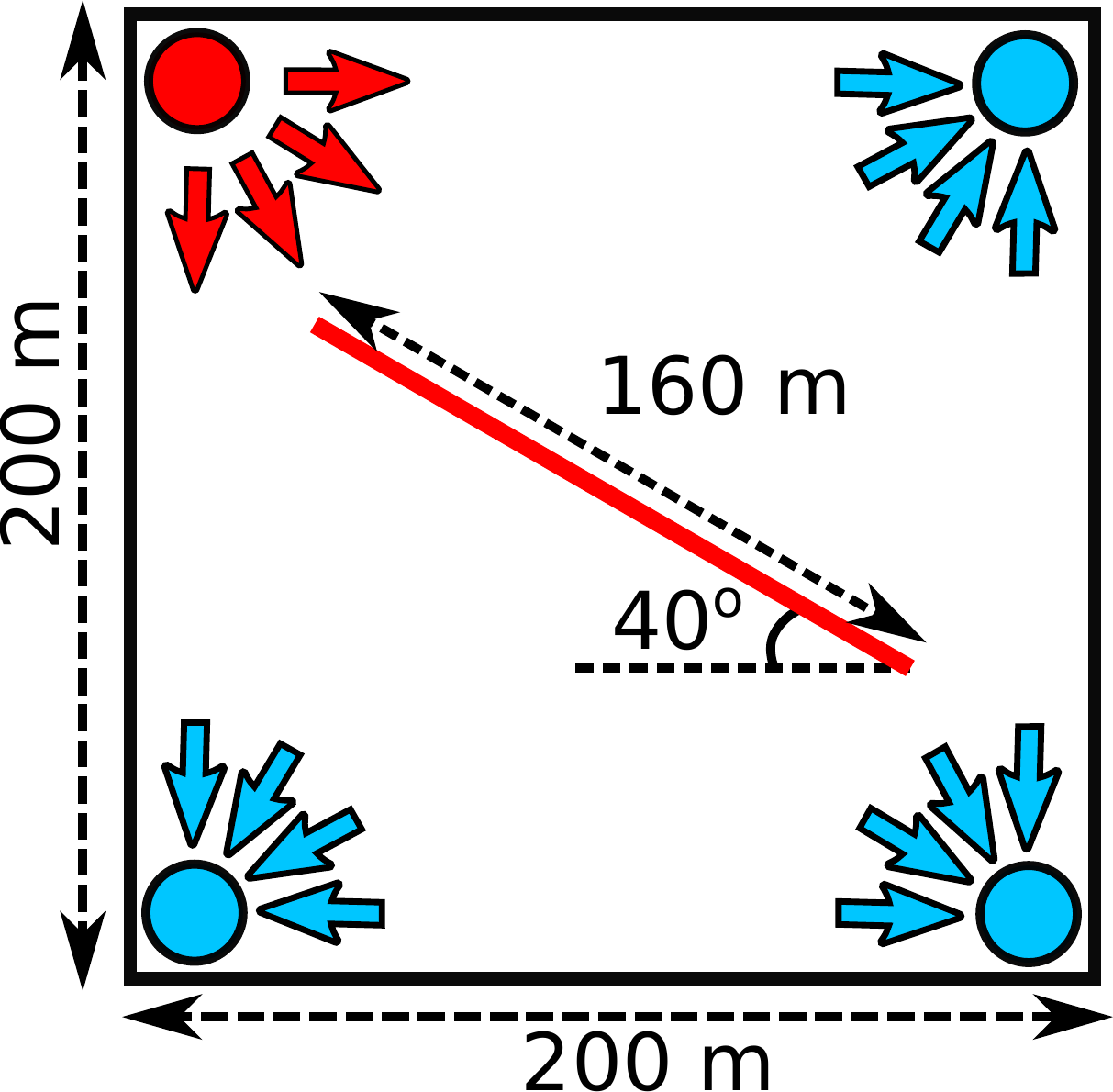}
    \caption{
      Domain schematics for the flow test problem.
      The red circle indicates an injection well.
      The three blue circles indicates production wells.
      The red straight line shows the fracture.
	}
	\label{fig:flow-domain}
\end{figure}

In this  section, we validate the  proposed approach for the  flow problem. The
2D-domain geometry  is the  same as  in the previous  tests (200  $\times$ 200
$\times$ 10 m$^3$).  A single fracture striking at 140$\degree$  to the x-axis
is located in the domain center. The flow parameters of the problem are listed
in Table  \ref{tab:parameters-flow-simulation}. Four  wellbores are  placed in
the corners of the domain as shown in Fig. \ref{fig:flow-domain}. The wellbore
in  the  left upper  corner  of  the domain  is  injecting  fluid at  constant
bottom-hole pressure  15 MPa. The rest  three wells are production  wells with
fixed bottom-hole pressure 5 MPa. The initial reservoir pressure is 10 MPa. We
apply no-flow condition on all boundaries.

To validate  the EDFM flow model  and investigate its spatial  convergence, we
compare the pressure in the fracture with  that obtained with the DFM model on
a fine  grid ($h =$  0.25 m) assumed as  the reference solution.  We performed
four runs of the EDFM model on  Cartesian grids with the element sizes: $h=$ 4
m, 2  m, 1 m, and  0.5 m. We used  the DFM model on  a hexahedral unstructured
grid with the elements of approximately equal sizes: 4 m, 2 m, 1 m, 0.5 m, and
0.25 m, the last considered the reference solution.

\begin{table}[H]
  \centering
  \begin{tabular}{lc}
    \hline
    Property                           &  Value   \\
    \hline
    Domain size [m$^3$]                & 200 $\times$ 200 $\times$ 10 \\
    Fracture length [m]                & 160 \\
    Fracture strike [$\degree$]        & 140 \\
    Fracture conductivity [mD$\cdot$m] & 20 \\
    Matrix permeability [mD]           & 10 \\
    Rock porosity [-]                  & 0.2 \\
    Initial reservoir pressure [MPa]   & 10 \\
    \hline
  \end{tabular}
  \caption{
    Geometrical and material parameters used in the flow test problem.
  }
  \label{tab:parameters-flow-simulation}
\end{table}

\begin{figure}[H]
    \centering
    \includegraphics[width=0.9\textwidth]{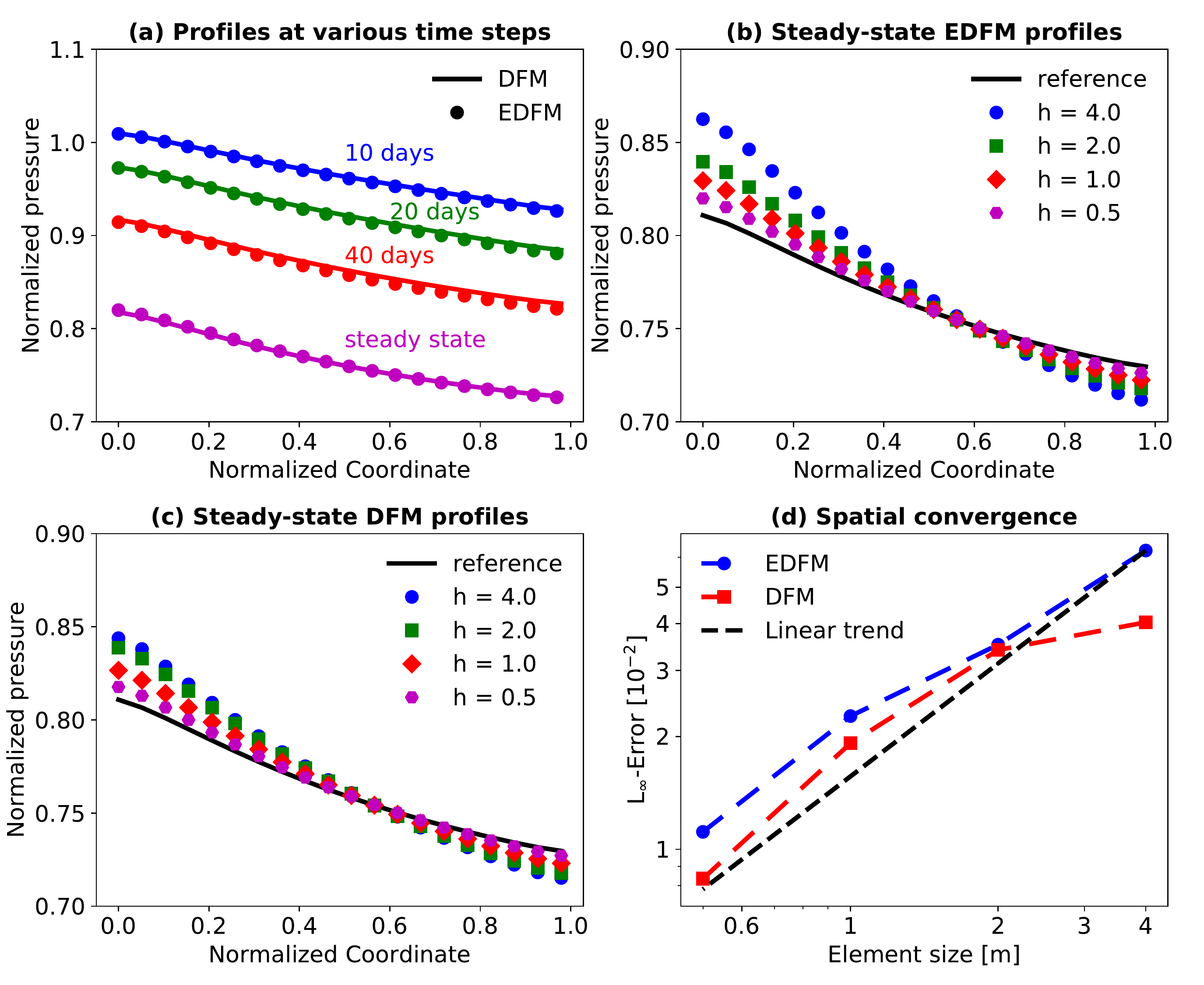}
    \caption{
      Simulation results for the flow problem.
      (a) Normalized fracture pressure as a function of coordinate along the fracture
      at various time instants given by DFM and EDFM models.
      (b)-(c) Steady-state fracture pressure given by the EDFM (b) and DFM (c)
      models at various grid resolutions.
      (d) Spatial convergence of the steady-state solutions:
      both DFM and EDFM models exhibit linear convergence.
	}
	\label{fig:flow-results}
\end{figure}

% THE RESULTS
%
The   simulation   results   for   the   flow   test   are   shown   in   Fig.
\ref{fig:flow-results}.  Fig.   \ref{fig:flow-results}a  shows   the  fracture
pressure as  a function of  the coordinate within the  fracture at 10,  20, 40
days, and  the steady-state solution.  The fracture pressure is  normalized by
the  initial reservoir  pressure,  and  the coordinate  is  normalized by  the
fracture length $l$.  The pressure in the fracture decreases  with time due to
the higher number of producing wells than injection wells. The results in Fig.
\ref{fig:flow-results}a  were obtained  from the  DFM and  EDFM models  on the
grids with the characteristic element size $h=$ 1.0 m. The DFM and EDFM models
provide relatively similar results at this refinement level.

Fig.  \ref{fig:flow-results}b-c   show  the  steady-state   fracture  pressure
profiles obtained at  various refinement levels. Evidently, both  DFM and EDFM
models overestimate fracture pressure closer to the injector and underestimate
it closer towards the producer on coarse grids.

Fig. \ref{fig:flow-results}d presents the spatial convergence of the results
given by DFM and EDFM models in the fluid test.
As evident from the figure, the EDFM and DFM models manifest very similar
convergence behavior with approximately the same slope.
We again point out that the results obtained from the DFM model, which may
explain a slightly lower error yielded by the DFM model.

\subsection{Coupled flow-mechanical test} \label{sec:results-coupled}
\begin{figure}[H]
  \centering
  \includegraphics[width=0.9\textwidth]{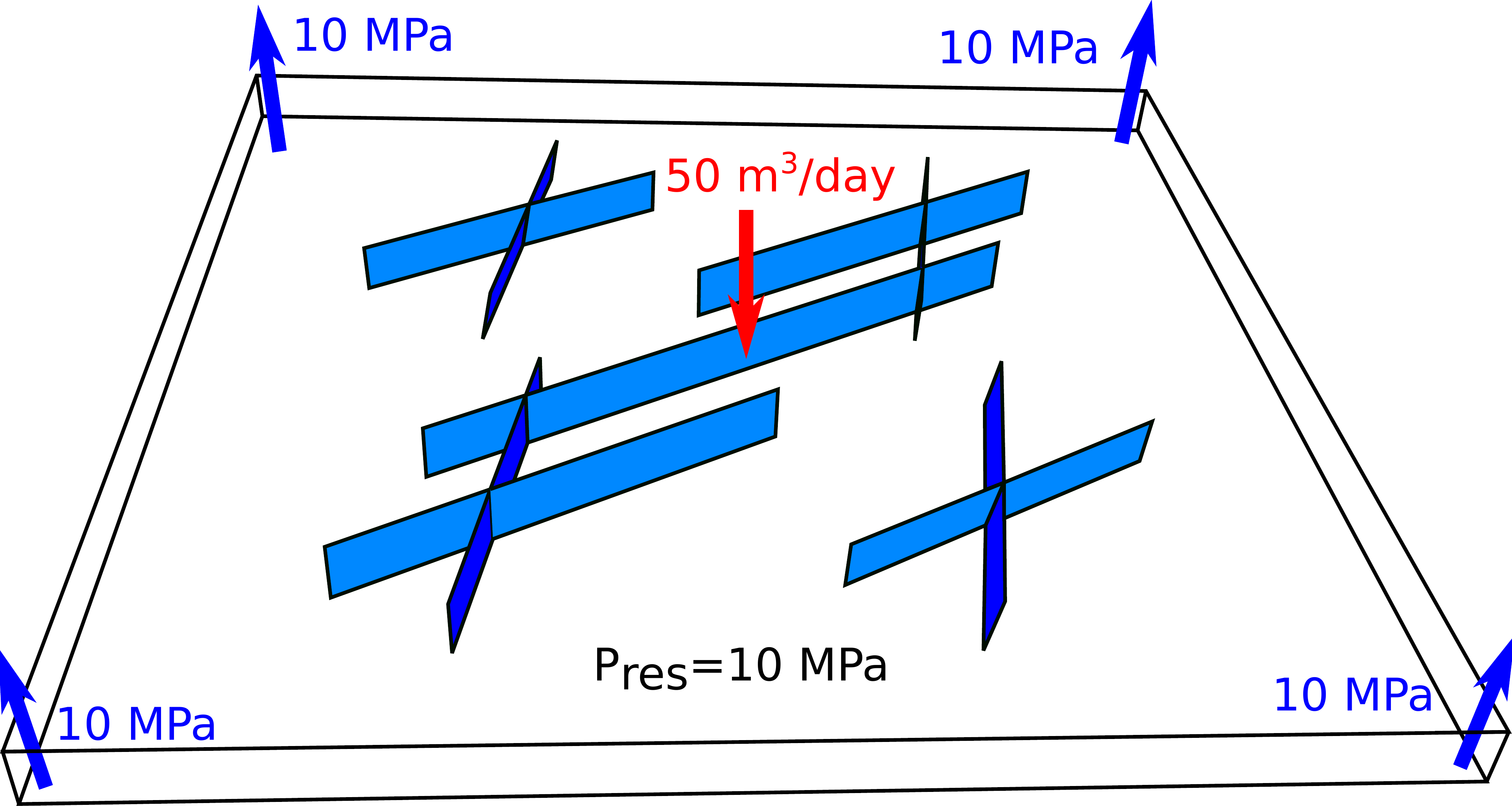}
  \caption{\label{fig:9frac-wells}
    Domain geometry for a coupled problem.
    The domain comprises nine fractures and five wells.
    The producing wells are shown in blue.
    The injection well is connected to the long
    fracture in the center of the domain and is shown in red.
  }
\end{figure}

In  this section  we present  a comparison  of the  DFM and  EDFM models  in a
coupled  3D  case.  Fig.~\ref{fig:9frac-wells}  shows a  reservoir  with  nine
variously-oriented fractures and five wellbores. The domain dimensions are the
same as  in the previous examples.  The initial reservoir pressure  is 10 MPa.
Four wells located  in the four corners  of the domain produce  fluid with the
bottom-hole pressure 10 MPa. Another well  at the center of the domain injects
fluid  with  constant rate  50  m$^3$/day  and  is  connected to  the  longest
fracture.   The  fracture   geometrical   properties  are   listed  in   Table
\ref{tab:parameters-9frac-simulation}.

\begin{figure}[H]
  \centering
  \includegraphics[width=\textwidth]{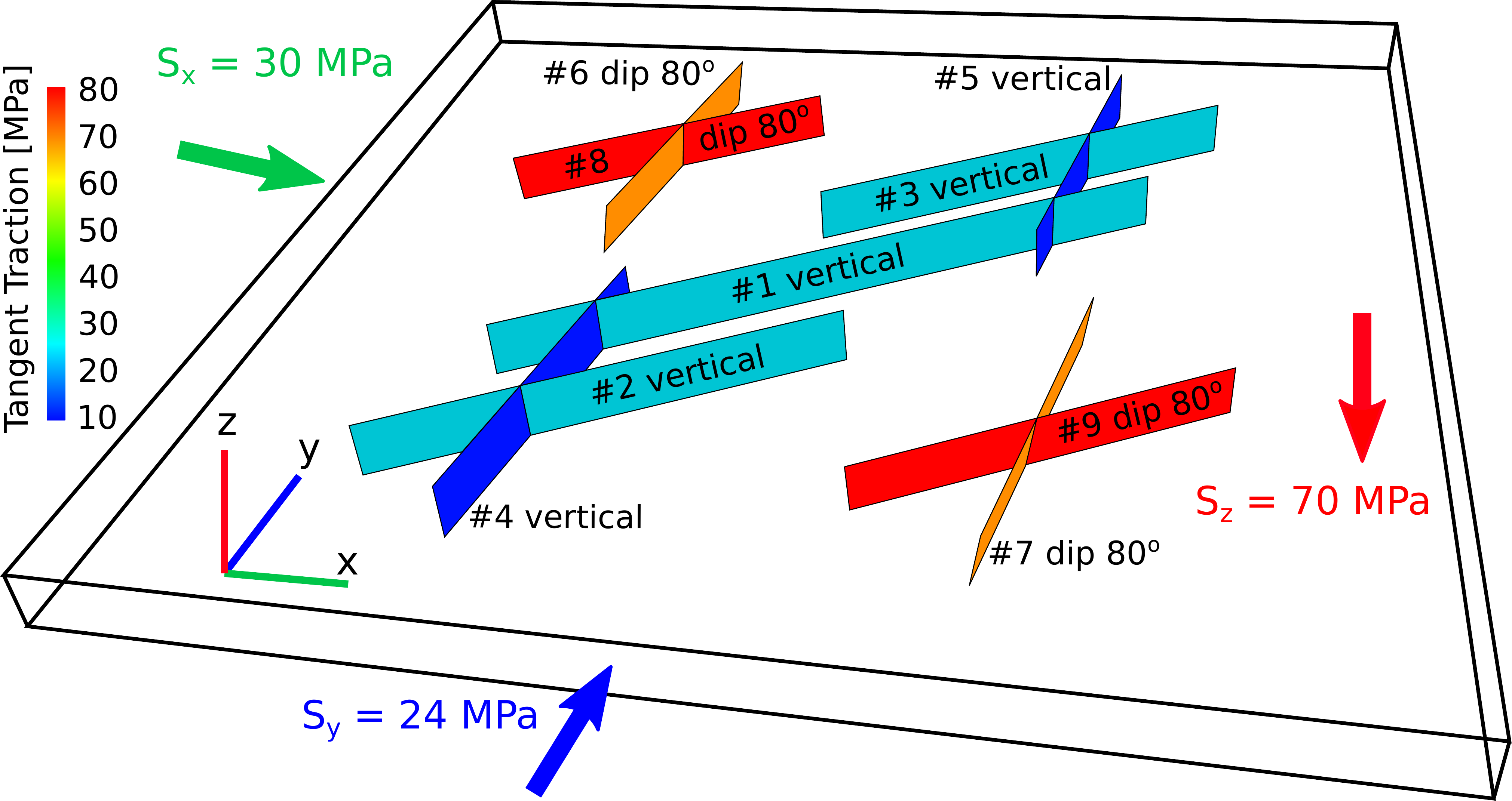}
  \caption{
    Initial stresses and tangential tractions on fracture surfaces.
    Vertical fractures \#4 and \#5 have the lowest initial tangent
    traction $t_t \approx$~10 MPa.
    Vertical fractures \#1, \#2, and \#3 have a medium
    initial tangent traction $t_t \approx$~25 MPa.
    Fractures \#6 and \#7 dip at 80$\degree$ and have a high
    initial tangent traction $t_t \approx$~70 MPa.
    Critically-oriented fractures \#8 and \#9 dip at 80$\degree$ and have the highest
    tangent traction $t_t \approx$~80 MPa.
  }\label{fig:9frac-stresses}
\end{figure}

The reservoir  is subjected  to anisotropic stresses  as follows:  the maximum
total horizontal  stress $S_{H max} = S_x$ = 30  MPa, the minimum total  horizontal stress
$S_{h min} = S_y$ =  24 MPa, and  the total vertical  stress $S_z$ =  70 MPa, as  shown in
Fig.~\ref{fig:9frac-stresses}.
The stress field  after the geomechanics initialization  procedure (an elastic
solution  at  time =  0)  results  in  various  tangent traction  $\bm{t}$  in
different   fractures.  Fractures   \#4  and   \#5  are   vertical  (have   no
vertical-stress component) and nearly  perpendicular to the maximum horizontal
stress $S_{H max}$.  Therefore, they have the lowest  initial tangent traction
$t_{\tau}$ $\approx$ 10 MPa. Fractures \#1, \#2, and \#3 are also vertical and
strike at 30$\degree$ to the direction  of the maximum horizontal stress $S_{H
max}$, which  causes a medium  initial tangent traction $t_{\tau}  \approx$ 26
MPa. Fractures  \#6 and \#7 dip  at 80$\degree$, which results  in the initial
tangent traction to  $t_{\tau}$ $\approx$ 70 MPa due to  the non-zero vertical
stress  component. Finally,  fractures \#8  and \#9  have the  highest tangent
traction $t_{\tau}$ $\approx$  80 MPa because of  the non-vertical orientation
(dip = 80$\degree$)  and the strike aligned closely to  the maximum horizontal
stress  $S_{H  max}$.   All  the  fractures  have   constant  conductivity  20
mD$\cdot$m.

We applied the presented DFM and EDFM models to simulate 60 days of injection.
The  computational domain  for  the  DFM simulation  was  discretized with  an
unstructured grid  with 59032 tetrahedrons. The  grid is fine at  the fracture
faces (2.2 m) and  coarse at the outer domain boundaries (10  m). The mesh has
four cells in the vertical direction  near the fractures. The EDFM simulations
utilized a  Cartesian grid  with 40804  hexahedrons. The  element size  in the
horizontal plane is 2 m. Same as the DFM grid, the EDFM grid has four cells in
the vertical (z) direction.

\begin{table}[H]
  \centering
  \begin{tabular}{llll}
    \hline
    Fracture \# &  Length [m] & Strike [$\degree$] & Dip [$\degree$]   \\
    \hline
    1           & 120         & 30                 & 90 \\
    2 and 3     & 80          & 30                 & 90 \\
    4 and 5     & 60          & 80                 & 90 \\
    6 and 7     & 60          & 80                 & 80 \\
    8 and 9     & 60          & 30                 & 80 \\
    \hline
  \end{tabular}
  \caption{
    Fracture geometrical parameters used in the coupled problem.
  }
  \label{tab:parameters-9frac-simulation}
\end{table}

\begin{figure}[H]
  \centering
  \includegraphics[width=\textwidth]{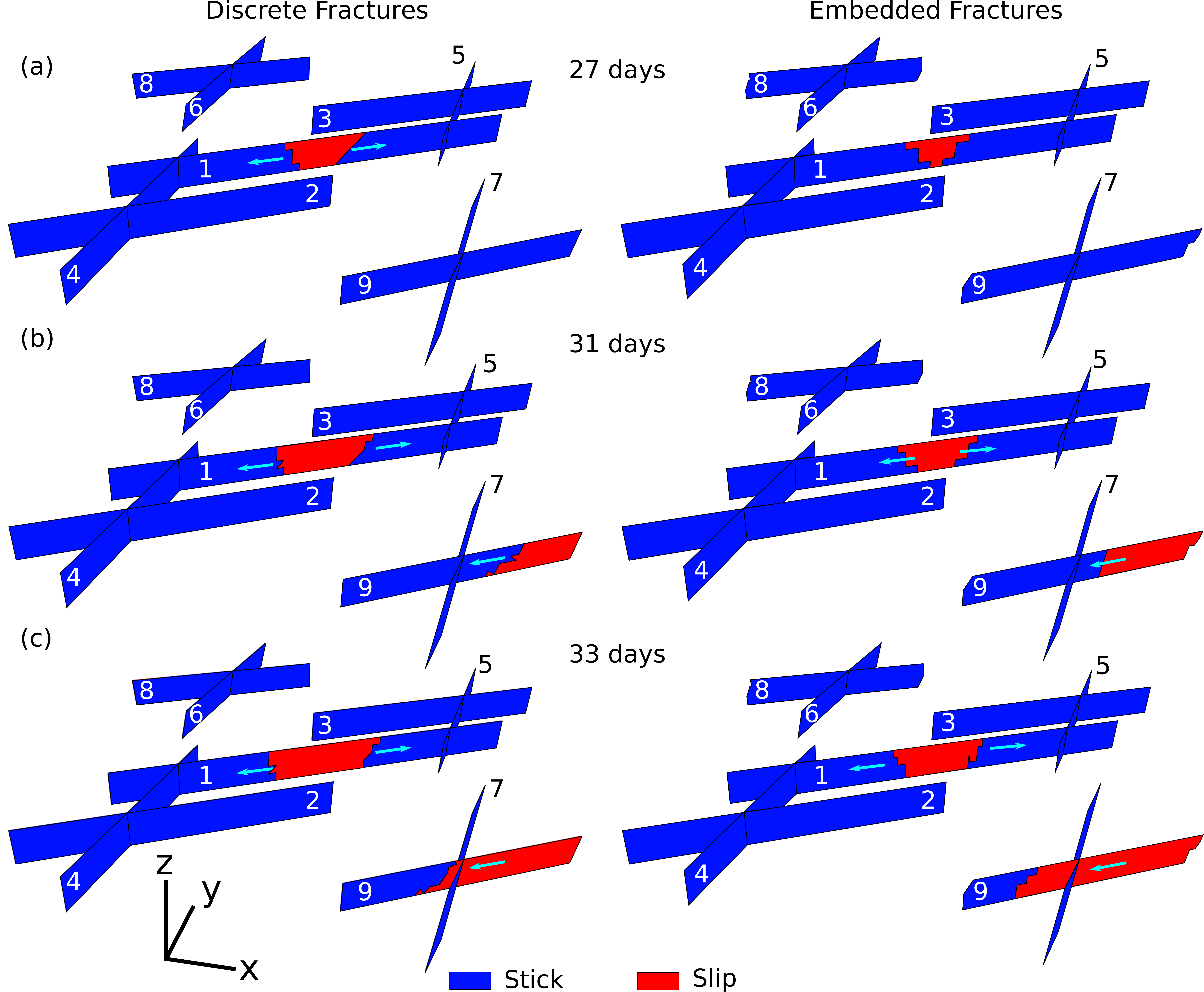}
  \caption{
    Development of the fracture states predicted by the DFM (left column)
    and EDFM (right column) methods.
    The figure shows which fracture elements are inactive (stick) or are
    active and exhibit slip or opening at various time steps.
    (a) At 27 days of injection the longest fracture \#1 (connected to the injector)
    is slipping with about 20\% of the area activated.
    The slipping region grows laterally in two directions from the fracture center.
    (b) At 31 days about 40\% of the critically-oriented fracture \#9 is slipping.
    The fracture activated at the farthest point in x direction, and the active
    region extends laterally.
    (c) After 33 days 5-15 \% of fracture \#8 is activated starting from the point
    with the lowest x-coordinate.
    Fracture \#9 is activated throughout 60-80 \% of its area.
    The activated area in the fracture \#1 is approximately the same as at 31 days.
  }\label{fig:9frac-statuses}
\end{figure}

The  process  of   induced  fracture  reactivation  is   illustrated  in  Fig.
\ref{fig:9frac-statuses}. At  approximately 15  days of injection  the longest
Fracture \#1 is activated  in the center (not shown in Figure).  At 27 days of
injection  about  20\%  of  the  area  of  Fracture  \#1  is  in  shear  (Fig.
\ref{fig:9frac-statuses})a. The  activated area is expanding  laterally in two
directions. The  DFM model predicts a  higher area of the  activated region in
Fracture \#1 at 27 days.
At 31  days of injection, a  larger portion ($\approx$ 30\%  area) of Fracture
\#1 is activated  (Fig. \ref{fig:9frac-stresses}b). At the  same time instant,
approximately  40 \%  area  of Fracture  \#9 is  slipping.  The activation  of
Fracture  \#9  commenced  at  approximately  19  days  (not  shown  in  Figure
\ref{fig:9frac-stresses}).
At 33 days  Fracture \#8 starts to slip  (Fig. \ref{fig:9frac-statuses}c). The
activated region  in Fracture  \#9 continues to  expand laterally  towards the
inactive end of the fracture. In  Fracture \#1, the activated region continues
to grow  bidirectionally.

\begin{figure}[H]
  \centering
  \includegraphics[width=\textwidth]{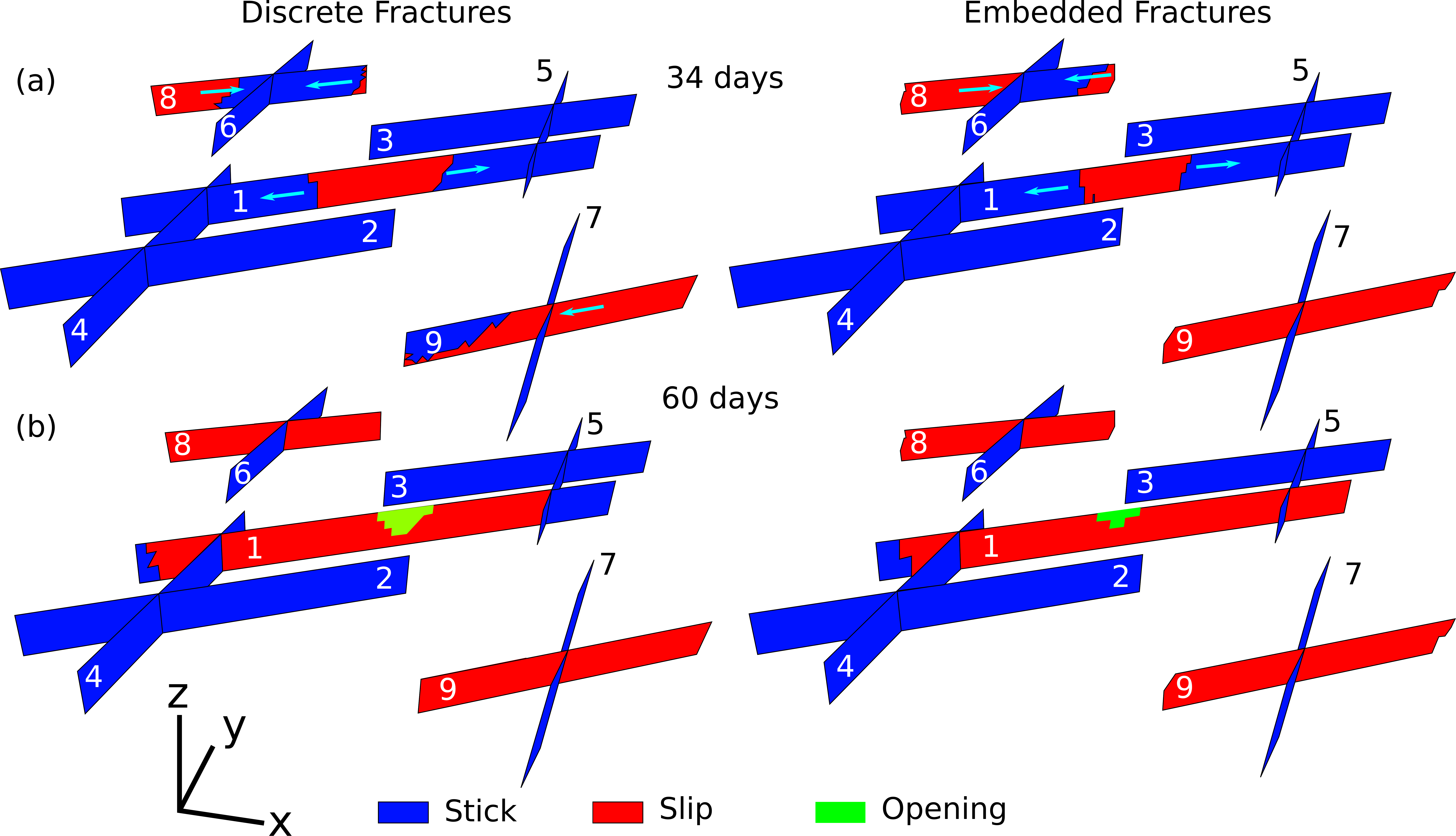}
  \caption{
    Development of the fracture states predicted by the DFM (left column)
    and EDFM (right column) methods.
    (a) Fracture \#8 is also activated on the another end at 34 days.
    From 50 to 70 \% of the area of fracture \#8 is slipping.
    (b) Fracture \#1 exhibits opening in the center at 60 days.
  }\label{fig:9frac-statuses-1}
\end{figure}

At 34 days, the opposite end of Fracture \#8 is activated in shear so that two
slipping  regions  within  the  fracture  are  expanding  towards  each  other
\ref{fig:9frac-statuses-1}a.  Following the  activation  of  the fractures  in
shear,  tensile  activation  of  Fracture  \#1  occurred.  After  60  days  of
injection,     a    small     portion    of     Fracture    \#1     is    open
(Fig.~\ref{fig:9frac-statuses}b.
Note that Fractures \#2-\#7 remained inactive during the injection period.
As shown in Fig. \ref{fig:9frac-statuses} and Fig. \ref{fig:9frac-statuses-1},
the DFM and EDFM models yield
qualitatively very similar behavior of fracture activation in this
complex scenario.

\begin{figure}[H]
  \centering
  \includegraphics[width=\textwidth]{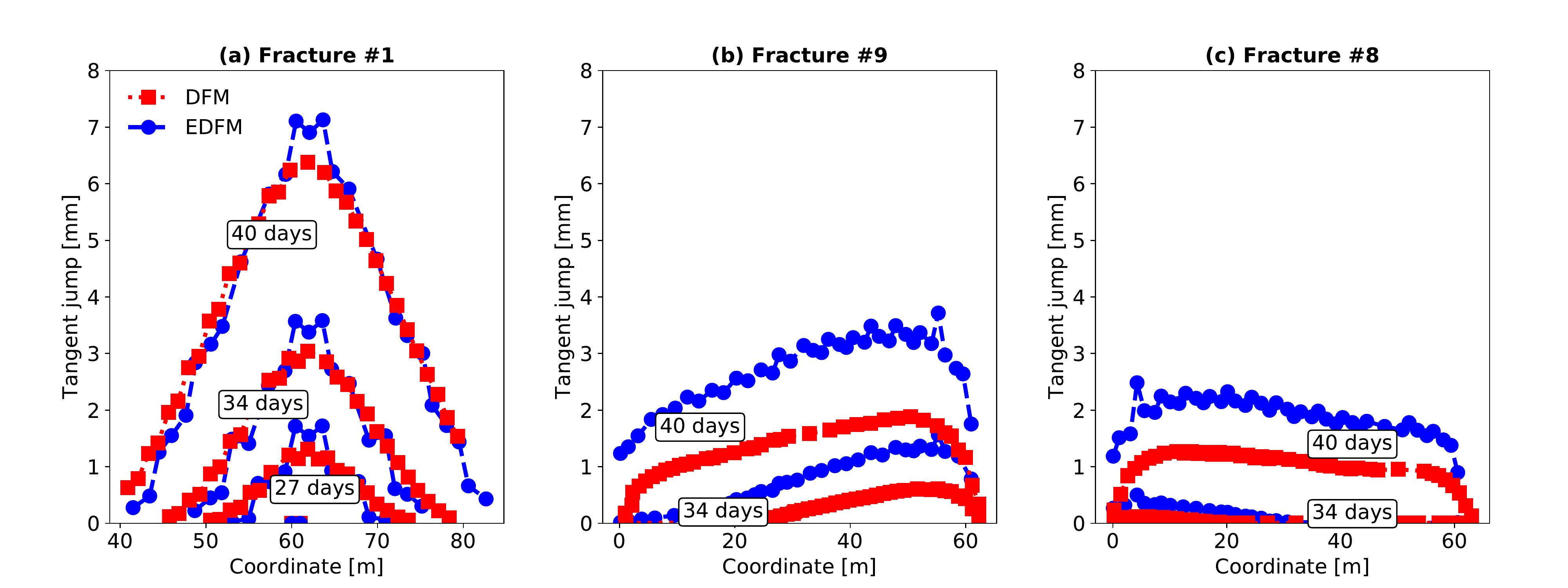}
  \caption{
    Fracture slip values as a functions of the coordinate at
    various time steps in
    (a) Fracture \#1.
    (b) Fracture \#9.
    (c) Fracture \#8.
  }\label{fig:9frac-displacement}
\end{figure}

We now compare the  fracture slip values predicted by the  DFM and EDFM models
within   the   fractures  \#1,   \#2,   and   \#3   at  various   time   steps
(Fig.~\ref{fig:9frac-displacement}).    The     jump    values     shown    in
Fig.~\ref{fig:9frac-displacement}  are obtained  along straight  lines at  the
vertical centers  of the  fractures. Fig.  \ref{fig:9frac-displacement}a shows
the  slip values  of  the longest  Fracture  \#1 at  27, 34,  and  40 days  of
injection as functions of the coordinate  along the fracture. The maximum slip
across    the    fracture   increases    with    time.    As   evident    from
Fig.~\ref{fig:9frac-statuses}, the active region  within the fracture expands,
and, therefore, the tangential jump within  the fracture is localized within a
larger region at late times than than  at early times. The DFM and EDFM models
yield similar values of the slip and similar activated areas for Fracture \#1.

Fig.~\ref{fig:9frac-displacement}b  shows the  tangential  jump values  within
Fracture \#9 at 34 and 40 days. We are not showing the curve for 27 days since
the  fracture was  completely  inactive  at that  time  instant. The  fracture
activated on  the right  (coordinate $\approx$ 60),  and the  activated region
expanded to the left. According to the  EDFM mode, at 34 days the fracture was
completely activated (Fig.~\ref{fig:9frac-statuses-1}),  whereas the DFM model
yields only approximately  three quarters of the fracture  activated. The DFM
model  also predicts  a lower  slip value  than the  EDFM model.  At 40  days,
Fracture \#9 is fully-activated according to both DFM and EDFM models.

Fig. \ref{fig:9frac-displacement}c shows the slip values in Fracture \#8 at 34
and  40 days.  Similarly to  Fracture \#9,  the EDFM  model predicts  a larger
activated area  in Fracture  \#8 at  34 days  than that  predicted by  the DFM
model. At 40 days both models predict the full activation of the fracture. The
EDFM model predicts  higher slip values than  the DFM model at both  34 and 40
days.

The activation of Fractures \#8 and \#9 is due to the increased fluid pressure
that reduces the fracture normal traction  and drives the fracture to slip. At
the same time, the fact that Fracture \#9 activated before Fracture \#8 is due
to  the  asymmetry of  the  mechanical  problem:  we fixed  the  corresponding
displacement  components  at  the  bottom, left,  and  front  boundaries,  and
assigned normal tractions on the opposing boundaries.

All in all, the presented hydro-mechanical simulations evidence the complexity
of  the fracture  behavior during  fluid injection.  Our DFM  and EDFM  models
provide  qualitatively  similar results  and  yield  similar dynamics  of  the
fracture activation.  The difference in the  tangential jump given by  the DFM
and EDFM  models are due to  slight differences in fluid  pressure and because
the  EDFM   model  overestimates   fracture  slip,   whereas  the   DFM  model
underestimates it as exemplified in previous sections.

% Nah these are results. we do not do interpretation and evaluation in results.
% \timur[All in all, the presented hydro-mechanical simulations show that the fractures
% behavior during water injection is a complex process.
% Due to excessive pressure and substantial stress changes, fractures can be activated and be
% in a slipping and opening modes.
% A robust and reliable hydro-mechanical simulator is required to solve such a challenging problem.
% EDFM and DFM methods enable us to solve the problem and demonstrate qualitatively similar results.]

% pressure does not reallty help understand anything
% \begin{figure}[H]
%   \centering
%   \includegraphics[width=\textwidth]{img/9frac-pressure.pdf}
%   \caption{
%     Fluid pressure values as a functions of the coordinate at
%     various time steps in
%     (a) Fracture \#1.
%     (a) Fracture \#9.
%     (a) Fracture \#8.
%   }\label{fig:9frac-displacement}
% \end{figure}

\section{Discussion}
This  section  evaluates  the  performance  of the  proposed  EDFM  model  and
discusses its limitations. We also provide  an insight into the modeling cases
where using EDFM or DFM is more preferable.

% \subsection{Limitations and evaluation}
\subsection{Limitations of the model}
%
% mechanics evaluation and comparison
%
All the  examples with slipping  fractures shown in  Section \ref{sec:results}
indicate that EDFM overestimates the slip. Moreover, the results evidence that
the tangential jumps given by EDFM  are always higher than those computed with
the DFM model.
This is presumably due to the locking  effect in the proximity of the fracture
tip  caused   by  the   continuous  slip  interpolation   in  the   DFM  model
\cite{borja2013plasticity}. The near-tip solution  can potentially be improved
by  assuming  non-constant  jump within  an  element  \cite{linder2007finite}.
Additionally,  using  EDFM  on  non-conforming  grids  results  in  non-smooth
displacement  profiles, which,  in  turn, may  cause  non-physical stress  and
traction values in active fractures. This  trait of the numerical model may be
undesirable   in   stress-sensitive   reservoirs,   especially   if   fracture
permeability     is      computed     as     a     function      of     stress
\cite{rutqvist2015fractured,shovkun2017coupled}.  In addition,  this also  may
cause convergence issues in fully-coupled models. Therefore, we suggest to use
the DFM approach in cases when a smooth jump is required.

The issue with non-realistic stresses in the SDA models can be circumvented by
enforcing  the continuity  of the  jump \cite{deb2017modeling}.  Requiring it,
however, results in a non-locality in  the numerical treatment of the jump. An
alternative way to  circumvent the deficiencies associated  with computing the
mechanics-dependent fracture permeability  with the presented SDA  model is to
assign    permeability     as    a     function    of     fracture    aperture
\cite{zimmerman1991lubrication,shovkun2019propagation}.     Since     fracture
aperture is widely  used to calculate the permeability  of open-mode fractures
with the lubrication theory \cite{lee2017iterative,shovkun2019propagation} and
given  the  capability  to  handle   arbitrary  fracture  geometries,  SDA  is
potentially  suitable for  modeling open-mode  fracture propagation.  A hybrid
formulation that makes  use of both SDA  and DFM methods for  various types of
fractures can potentially be an  optimal approach for coupled hydro-mechanical
modelling of fractured reservoirs.

% \subsection{Limitations and evaluation my version}
% %
% According to our results in Section \ref{sec:results}
% EDFM approach overestimates the shear slip.
% Moreover, the results evidence that the tangential jumps given by EDFM are
% higher than those computed with the DFM model.
% This is presumably due to the locking effect in the proximity of the fracture tip
% caused by the continuous slip interpolation in the DFM model \cite{borja2013plasticity}.
% %
% Additionally, the EDFM approach produces non-smooth
% displacement profiles especially on non-conforming grids. This circumstance can be important for modeling stress-sensitive reservoirs,
% especially if fracture conductivity is computed as a function of stress
% \cite{rutqvist2015fractured,shovkun2017coupled} and aperture \cite{lee2017iterative,shovkun2019propagation} .
% Our simulations show that this problem can be partially resolved by grid refinement.
% Thus, the computational grid resolution should be selected with care.

% \subsection{Fracture intersection}
\subsection{SDA formulation for intersecting fractures}
In this section  we outline the basic idea behind  the mechanical treatment of
fracture intersection in our SDA model.  When more than one embedded fracture
are located in  a grid element, the contact mechanics is  described with the following
system:
\begin{subequations} \label{eq:sda-frac-intersection}
  \begin{align}
    & \dot{\bm{\sigma}} = \varmathbb{C} : \left[ \nabla^s \dot{\bm{u_c}} -
          \sum_{ i=1 }^{N_F} \left( [\dot{\bm{u}}]_i \otimes \nabla f_i \right)^s \right]  \\
    & \left[\dot{\bm{u}}\right]_i = \lambda_{\delta_i} \frac{\partial G_i}{\partial \bm{t}_i} ,\quad i=1...N_F\\
    & \dot{q_i} = -\lambda_{\delta_i} {H_{\delta_i}} \frac{\partial G_i}{\partial q_i} ,\quad i=1...N_F \\
    & \lambda_{\delta_i} F_i(\bm{t}_i, q_i) = 0 ,\quad i=1...N_F
  \end{align}
\end{subequations}
where
$N_F$  is the number of embedded fractures in the cell,
$\left[ \bm{u} \right]_i$ are the fracture jump vectors,
$\lambda_i$ are fracture plastic multipliers,
$q_i$ are the fracture internal state variables,
$H_{\delta_i}$ are the fracture hardening moduli,
$\bm{t}_i$ is the fracture traction,
and $F_i$ and $G_i$ are the flow rules plastic potentials, respectively.
The  system   \eqref{eq:sda-frac-intersection}  is  the  direct   analogue  to
\eqref{eq:sda-system}. All  other equations and the  solution procedure remain
the same as in the case of  a single fracture. Therefore, adding a fracture to
the    system   consists    of    (1)    imposing   independent    constraints
\eqref{eq:sda-frac-intersection}b-c  on  variables  related  to  a  particular
fracture and (2) using the superposition of jumps to determine the stress (Eq.
\ref{eq:sda-frac-intersection}a).  Thus,  coupling several  EDFM fractures
within a cell occurs via the effective stress~$\bm{\sigma}$.

This      strategy     has      two     drawbacks.      First,     formulation
\eqref{eq:sda-frac-intersection}  permits  the   maximum  of  two  embedded
fractures per grid element. Applying this  approach to more than two fractures
per element results in an over-determined  system.

Second,  each combination  of  fracture  states (e.g.  the  first fracture  is
slipping and  the second  fracture is opening)  requires a  special treatment.
This      treatment     consists      in     relaxing      the     constraints
\eqref{eq:shear-flow-and-potential}-~\eqref{eq:sda-stick}    to   avoid    the
over-determination of the system~\eqref{eq:sda-frac-intersection}. Future work
is needed  to develop  a robust procedure  for treating  fracture intersection
with various statuses.

% \timur[This section is not complete.
% The system 29 is too general and doesn't tell how to solve the problem.
% 1. We can start with a comment that we can intersect 2 fractures (we reaaly do only 2 fractures).
% 2. Then provide a system.
% 3. Then talk about drawbacks.
% 4. We can also point out that this approach works just fine for stick-"slip or open"
%intersections as shown in fig 13 and fig 14.
% 5. And conclude that this is an ongoing work. ]

\subsection{Flow modeling} \label{flow-limitations}
The    approach    to    modeling     embedded    fractures    discussed    in
Section~\ref{sec:frac-flow},  is a  simplified way  of describing  fluid flow.
While providing  a relatively  accurate approximation for  single-phase fluids
for  highly-permeable  fractures,  the   traditional  EDFM  approach  has  two
important  limitations. First,  standard EDFM  cannot simulate  fractures with
permeability lower than that of the rock matrix \cite{flemisch2018benchmarks}.
Second,  the  standard EDFM  cannot  accurately  capture cross-fracture  flow,
sometimes  referred to  as fracture  sweeping \cite{karvounis2013simulations}.
This effect becomes particularly significant when solving transport equations.
Two solutions  have been suggested  to remove these limitations.  Karvounis in
proposed to modify the computation  of the matrix-fracture flow $q_{MF}$ terms
based  on the  direction of  the cell  fluxes \cite{karvounis2013simulations}.
Projection  EDFM (pEDFM)  is  another  approach that  modifies  the number  of
connections in an approximation pattern \cite{ctene2017projection}. The latter
approach can be easily implemented in the connection-based flow simulator as a
part of the preprocessing stage \cite{KarimiFard2004},

\section{Conclusion}

% overall
%
% The difficulties  associated with  aligning computational grid  with fractures
% suggest Embedded Discrete  Fracture Method (EDFM) as an  potential strategy to
% model multiple arbitrarily-oriented fractures.
In this paper we presented a numerical  model  that  uses  the  Strong
Discontinuity  Approach  (SDA)  for
geomechanics and  the Embedded Fracture  (EDFM) approach for fluid  flow. This
study  validates  the use  of  a  coupled  EDFM  model in  relatively  complex
geological settings.

% mech convergence
%
A series of  mechanical tests was performed to compare  the performance of the
Discrete Fracture Model  (DFM) and the proposed EDFM model.  Both DFM and EDFM
manifest asymptotic super-linear convergence on conforming grids when modeling
slipping  and  opening fractures.  On  non-conforming  grids, the  EDFM  model
manifests  a  range  of  convergence   behaviors  depending  on  the  fracture
orientation with respect to the  computational grids. On average, however, the
EDFM model  retains the  super-linear convergence trend  for both  opening and
slipping fractures.

% mech conceptual results
%
Numerical  simulation evidences  discontinuous slip  and aperture  within EDFM
fractures. On very coarse grids, the  absence of jump continuity may result in
non-physical jump  profiles. Based  on the simulations  we conclude  that EDFM
overestimates fracture slip, whereas DFM underestimates it.

We presented a simplified EDFM approach to simulate single-phase fluid flow in
fractured porous media. The presented model provides accurate results that are
very close to those given by the  DFM model. We demonstrated that both methods
have similar linear convergence behavior.

Finally, we applied the DFM and EDFM  methods to simulate a complex 3D coupled
hydro-mechanical  problem.  We  considered  a fluid  injection  scenario  that
resulted in  natural fractures  activation induced  by excessive  pressure and
stress alterations.

Our results indicate  that EDFM and DFM methods  provide qualitatively similar
results.  Based on  multiple numerical  tests, we  conclude that  EDFM can  be
considered an alternative approach to model highly fractured reservoirs.

\section*{Acknowledgments}
We thank the Reservoir Simulation Industrial Affiliates
Consortium at Stanford University (SUPRI-B)  for funding this research.

\bibliography{bibliography}{}
\bibliographystyle{apalike}

\end{document}